# The Evolution of IJHCS and CHI: A Quantitative Analysis

Andrea Mannocci, Francesco Osborne and Enrico Motta
Knowledge Media Institute
The Open University
*{andrea.mannocci, francesco.osborne, enrico.motta}@open.ac.uk*

## Abstract

In this paper we focus on the International Journal of Human-Computer Studies (IJHCS) as a domain of analysis, to gain insights about its evolution in the past 50 years and what this evolution tells us about the research landscape associated with the journal. To this purpose we use techniques from the field of Science of Science and analyse the relevant scholarly data to identify a variety of phenomena, including significant geopolitical patterns, the key trends that emerge from a topic-centric analysis, and the insights that can be drawn from an analysis of citation data. Because the area of Human-Computer Interaction (HCI) has always been a central focus for IJHCS, we also include in the analysis the CHI conference, which is the premiere scientific venue in HCI. Analysing both venues provides more data points to our study and allows us to consider two alternative viewpoints on the evolution of HCI research.

## 1 Introduction

This paper is situated in the context of the Special Issue celebrating the 50th anniversary of the International Journal of Human-Computer Studies (IJHCS). As indicated in the call for papers, the Special Issue is concerned with landscape papers exploring the evolution of research in areas relevant to IJHCS. Within this context, it is interesting to focus on the journal itself as a domain of analysis, to gain insights about how it has evolved in the past 50 years and what this evolution tells us about the research landscape associated with the journal.

Hence, in this paper we present an analysis of the evolution of the journal as evidenced by the associated scholarly data. In particular, we investigate how IJHCS, as well as its predecessor until 1993, the International Journal of Man-Machine Studies[1], have evolved over the years in terms of demographics, citation patterns, evolution of research areas, and other parameters. This work is therefore situated within the field of *Science of Science* (SciSci), the discipline that studies the interactions among scientific agents [1] with the aim of gaining a better understanding of the research landscape and ultimately using these insights to accelerate scientific progress.

[1] For the sake of simplicity, in the following we use IJHCS to indicate both journals.

In order to provide more data points to the analysis, in addition to studying the data associated with IJHCS publications, we also consider the scholarly data associated with the Conference on Human Factors in Computing Systems (CHI), which is the premiere scientific venue in the area of Human-Computer Interaction (HCI). While the domains of CHI and IJHCS do not necessarily coincide entirely (in particular, IJHCS is an interdisciplinary journal with a broad scope that goes beyond HCI to include much influential research in knowledge based systems, knowledge acquisition, and ontology engineering), HCI has always been a central focus for IJHCS and, especially in the past decade, it is the case that the journal has focused more and more on its core area of innovative interactive systems. Hence, it seems to us that it is a natural choice to include CHI in our analysis, to provide an alternative viewpoint over the evolution of HCI research in the past few decades.

In what follows we will analyse the two scientific venues throughout their entire history: since 1982 in the case of CHI, since 1969 in the case of IJHCS.

# 2 Methodology

For the sake of rigour and reproducibility, we have followed a well-defined methodology and made available all relevant data and code[2]. Specifically, the analysis reported in this paper is structured according to the following steps:

- *Research question definition*, in which we define the research questions that we aim to address in this study;
- *Generation of the datasets*, in which we extract the relevant datasets from *Microsoft Academic Graph* [2];
- *Scientometric analysis*, in which we analyse the bibliographic metadata of the research articles published, cited by, and citing IJHCS and CHI, and produce several analytics to address the research questions;
- *Geopolitical analysis*, in which we use the framework introduced in [3] to highlight the key geopolitical trends;
- *Research topic analysis*, in which we classify IJHCS and CHI publications according to the research topics drawn from the *Computer Science Ontology* [4] and identify the most significant topic trends.

In what follows, we will describe the various steps of the process.

## 2.1 Research Questions

As already pointed out, we are interested in analysing the evolution of IJHCS and CHI with respect to different perspectives, concerned with spatial scientometrics [5], citation flows, and topic trends. Crucially, we are not concerned with individual authors or publications – e.g., most published author or most cited paper, and we focus instead on the analysis of trends defined at a level of granularity that is 'macro' enough to provide us with useful insights about the features of the research space, whether this concerns key topic trends or geopolitical aspects.

[2] Data and code of our analysis, https://github.com/andremann/IJHCS-special-issue. The repository is also available on Zenodo for long term preservation, https://doi.org/10.5281/zenodo.2671681.

To this purpose, we consider the following research questions:

1. Are there significant differences that emerge from analysing the research dynamics associated with IJHCS and CHI?
2. What are the insights that emerge from an analysis of the citation data associated with IJHCS and CHI, including both papers citing and being cited by IJHCS and CHI publications?
3. What are the most significant geopolitical patterns emerging from analysing authors' affiliations?
4. What are the main research trends that emerge from a topic-centric analysis of the data?

## 2.2 Generation of the Datasets

We performed our analysis using data drawn from Microsoft Academic Graph[3] (MAG), which is a pan-publisher, multidisciplinary scholarly dataset produced, maintained, and delivered by Microsoft Research [2]. At the time of writing, MAG is provided as a blob of TSV files via Microsoft Azure Storage (AS) and features an ODC-BY[4] licence, an aspect that is essential to ensure transparency and reproducibility of the results shown in this work. In addition, MAG provides, whenever possible, key data often not available in other publicly accessible datasets (e.g. Crossref[5], DBLP[6]), including authors' and affiliations' identifiers [6], which are required to address some of our research questions.

The MAG dataset provided via Microsoft Azure Storage was transferred to the *Big Data Cluster* available here at The Open University [7]. The six datasets used in our analysis, which include IJHCS and CHI publications plus all papers cited by or citing them, were then extracted via *Spark* and saved as *TSV files*. The code for performing data extraction as well as the actual datasets can be found at https://github.com/andremann/IJHCS-special-issue/tree/master/src/spark. For the sake of convenience, in what follows we will refer to the resulting set of scholarly data as the *IJHCS+CHI dataset*.

Each paper in the resulting dataset is characterised in terms of the following properties: title, abstract, relevant topics, authors and their affiliations, author order, venue, year, and various identifiers, which include the ID and DOI of the paper and the IDs of all cited papers. Unfortunately, some publications in MAG may lack some of these properties – e.g., IDs of author affiliations and cited papers are sometimes missing. In Table 1 we show the size of the datasets.

[3] Microsoft Academic, https://aka.ms/msracad.

[4] Open Data Commons Attribution Licence (ODC-BY) v1.0, https://opendatacommons.org/licenses/by/1.0/.

[5] Crossref API, https://github.com/CrossRef/rest-api-doc.

[6] DBLP, https://dblp.uni-trier.de.

| | Accepted papers | Citing papers | Cited papers |
|---|---|---|---|
| **IJHCS** | 3,255 | 147,307 | 91,158 |
| **CHI** | 15,738 | 423,500 | 244,301 |

*Table 1 - Size of the IJHCS+CHI dataset.*

## 2.3 Scientometric Analysis

In this initial analysis, we identified the key institutions and studied the relationships between IJHCS, CHI, and the other major venues cited by and citing them. In order to detect the most prominent institutions, we used the author affiliations in the metadata, hereafter indicated as *contributions*, in accordance with the approach introduced in [3]. For example, if paper *p* is authored by authors $a_1$ and $a_2$, two distinct contributions are taken into account, one for each author. To disambiguate authors and institutions, we relied on the author and GRID[7] identifiers included in the data.

## 2.4 Geopolitical Analysis

We examined the contributions to IJHCS and CHI from different countries and ranked them by number of publications. We then repeated this analysis for the set of papers cited by and citing publications in IJHCS and CHI. We also tried to assess to what extent the research landscape associated with IJHCS and CHI is *open* or not, by measuring the variation of country rankings over the years, as suggested in [3]. To this end, we assessed the correlation of country rankings in subsequent years according to *Spearman's rank non-parametric correlation coefficient* (Spearman's rho) [8], which is a non-parametric measure of rank correlation in the range [-1, 1]. In order to highlight the key players in the field, we also identified the countries capable of producing papers without foreign collaborations – i.e. the authors of these papers are all from the same country.

To further characterise the relationship between countries and venues, we also considered the concept of *knowledge debit*, which is defined as follows: when a paper $p_v$ published in a venue *v* cites a paper $p_i$, the countries that participate in $p_i$ build up their *credit* towards the venue. Conversely, when a paper $p_v$ published in *v* is cited by a paper $p_j$, the countries that participate in $p_j$ build up their *debit* towards the venue. Essentially, the idea is that some countries may build a sizeable 'debit' towards a venue – i.e., they cite it a lot but the publications in the venue cite papers from this country much less. This situation would indicate an imbalance which may be related to the inability of these countries either to publish at all in the venue in question or, if they are published, to make an impact. Formally, we define *knowledge debit* as follows:

$$knowledge_{debit}(c,v) = \frac{\#\ contributions_{citing}(c,v)}{\#\ contributions_{cited_by}(c,v)}$$

Finally, we analysed the affiliations that were never represented by a first author, which is traditionally considered the key contributor [9], with the aim of identifying institutions and countries that have participated to IJHCS and CHI, but that (according to most traditions, at least) were not the main lead on data collection, analysis or write-up.

[7] Global Research Identifier Database (GRID), https://grid.ac.

## 2.5 Research Topics Analysis

The analysis of the research topics reported in this paper follows the *Expert-Driven Automatic Methodology* (EDAM) [10], a methodology designed for reducing the amount of manual tedious tasks involved in *mapping studies* [11]. Mapping studies are rigorous and reproducible methodologies to build classification schemes and analyse frequencies of publications for categories within the scheme, with the ultimate aim of analysing trends in the research literature. While in traditional mapping studies [11, 12], the authors are required to manually analyse a large number of papers, EDAM automates the classification process by exploiting an unsupervised approach to characterizing publications according to the research topics in a domain ontology. It is therefore particularly appropriate for analysing large datasets of research papers and producing analytics regarding topics trends.

We adopted a simple version of the EDAM methodology which follows these steps:

1. *Selection of the Domain Ontology*, in which we adopted the Computer Science Ontology as a suitable classification for systematising the relevant research areas;
2. *Classification of Primary Studies*[8], in which we classified the articles published in IJHCS and CHI according to research topics drawn from CSO;
3. *Data Synthesis*, in which we produced the various analytics regarding the popularity of the research topics over the years.

In the next subsections, we describe the various steps of the process.

### 2.5.1 The Computer Science Ontology

The Computer Science Ontology (CSO)[9] [4] is a large-scale, automatically generated ontology of research areas, which includes about 14K topics and 162K semantic relationships. CSO was produced by applying the Klink-2 algorithm on a large corpus of publications [13] and has been used to support a range of applications and methods for community detection, trend forecasting, and paper classification – see [4] for a comprehensive review of the application of CSO to a variety of tasks. It was also adopted by two mapping studies in the fields of Semantic Web [14] and Software Architecture [10].

CSO presents several characteristics that make it particular appropriate to support our study. In the first instance, it is the most complete representation of research topics in Computer Science, about an order of magnitude larger than other current characterisations of this field, such as the ACM Classification[10], in terms of both number of topics and relationships. For instance, CSO currently includes 684 sub-topics of HCI, while the ACM Classification only contains 36. In addition, the CSO team recently released the *CSO classifier*, a tool for automatically classifying publications which was shown to generate excellent results [10, 15]. Finally, CSO is licensed under a Creative Commons Attribution 4.0 International License (CC BY 4.0)[11], which facilitates the reproducibility of our work.

[8] The term “primary studies” is used in mapping studies to refer to the body of literature under examination.

[9] CSO Ontology, cso.kmi.open.ac.uk.

[10] ACM Classification, https://www.acm.org/publications/class-2012.

### 2.5.2 Automatic Classification of Research Papers with the CSO Classifier

We annotated the papers in the IJHCS+CHI dataset by means of the CSO Classifier [15, 16], an unsupervised classifier that takes as input the metadata of a research article (title, abstract, and keywords), and returns a selection of the relevant research areas drawn from CSO. A first version of the CSO classifier has been in use since 2016 as a component of the Smart Topic Miner (STM) [17], the tool adopted by Springer Nature to annotate proceedings in the field of Computer Science, and the Smart Book Recommender (SBR) [18], an ontology-based recommender system for editorial products. This same approach was also used by several research prototypes that exploited the resulting topic representation for predicting technologies [19], detecting research communities [20], and forecasting research topics [21]. In particular, it was recently adopted by an implementation of the EDAM methodology in the field of Software Architectures [10], reporting a performance in classifying papers not significantly different from that exhibited by six senior researchers (p=0.77). In the current study, we use the version presented in [16], which is a slightly improved version of the classifier originally used in STM [17]. The most recent version of the CSO Classifier, which uses word embeddings to identify additional topics, is described in Salatino et al. [15].

The version of the CSO classifier adopted for our analysis operates accordingly to the following steps. First, it removes English stop words from the input text and extracts unigrams, bigrams, and trigrams. Then, for each n-gram, it computes the Levenshtein similarity with the labels of the topics in CSO. Research topics having a similarity with an n-gram, which is equal or higher than a threshold (this is set empirically to 0.94, as in [16]), are identified as relevant. In order to further enrich this set of topics, the CSO Classifier infers also their super topics by exploiting the *superTopicOf* relationships within CSO. For instance, given the topic "Neural Networks", it will also infer its super-topics, such as "Machine Learning" and "Artificial Intelligence". Finally, it uses the *relatedEquivalent* relationships in CSO to tidy up redundant concepts. For instance, it would detect that Ontology Matching and Ontology Mapping are equivalent topics, i.e., synonyms, and keep only the topics suggested by the *primaryLabel* relationship in CSO, in this case Ontology Matching.

The CSO Classifier was implemented in Python and the open-source codebase is available at https://github.com/angelosalatino/cso-classifier. An online demo can be accessed at https://cso.kmi.open.ac.uk/classify.

### 2.5.3 Data Synthesis

The average paper was associated with 13.9 topics in both IJHCS and CHI, including topics inferred through *skos:broaderGeneric* and *relatedEquivalent* relationships. Interestingly, when focusing on the last ten years (2009-2018), the average IJHCS paper covers a slightly larger number of topics (16.6) than the average CHI paper (14.2). The resulting CSV files reporting the evolution of topics were analysed to detect the most prominent topics in various timeframes and the most significant trends in the last ten years.

[11] CC BY 4.0 International License, https://creativecommons.org/licenses/by/4.0.

We paid particular attention to the last decade, 2009-2018, and we focused on the topics that experienced a steeper improvement in term of publications in CHI and IJHCS during this period. Since we wanted to consider both major and minor topics, we grouped the topics according to their magnitude. For IJHCS, we split the topics in two sets including respectively those with at least 10 publications in 2018 and those with less than 10 but more than 5 publications in the same year. For CHI, which publishes more papers than IJHCS, we split them in four groups containing the topics associated with at least 60, 20, 10, and 5 publications in 2018. We then ordered these topics according to the ratio between their number of publications in 2009 and in 2018. Finally, we reviewed the resulting lists with domain experts, discarded redundant topics, and selected 10 topics from each group. In order to analyse the contributions of specific countries to the main research topics, we also produced distributions of countries analogous to those described in Section 2.4, but considering only the publications associated to each topic.

# 3 Results

## 3.1 Scientometric Analysis

Figure 1 reports the number of papers accepted in IIJHCS (a) and CHI (b). As shown in the figure, the number of publications in IJHCS has remained relatively stable over the years, while the CHI conference has grown massively, now publishing over 1,200 papers per year (including extended abstracts, symposia, and workshops).

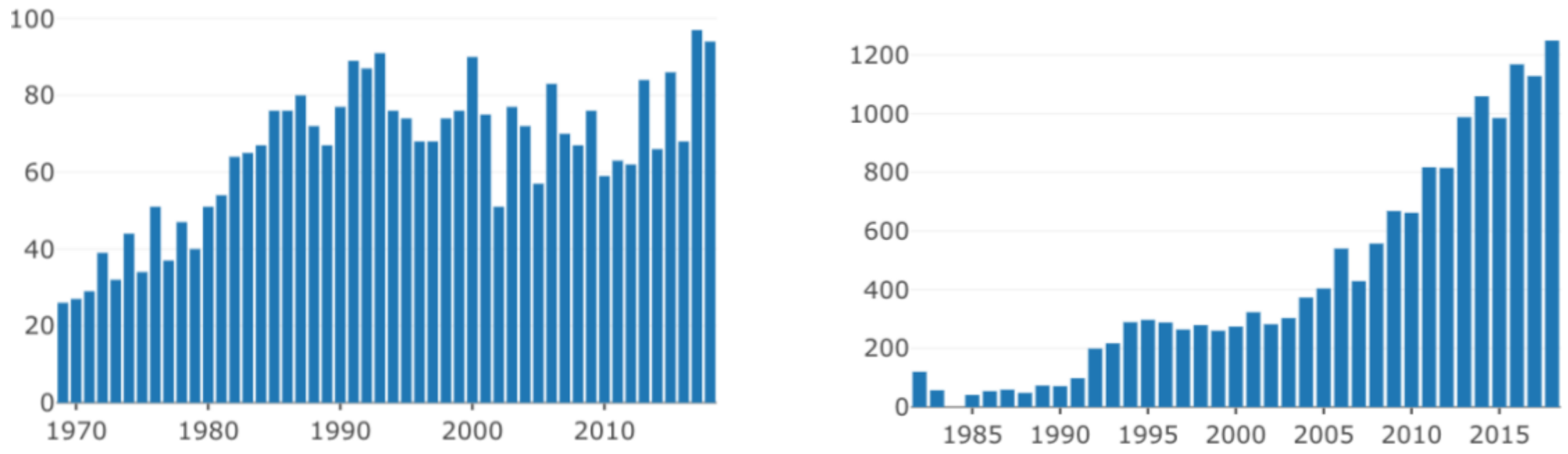

a) Papers published in IJHCS

b) Papers published in CHI

*Figure 1 - Number of papers accepted in IJHCS (a) and CHI (b). When comparing IJHCS and CHI, please consider that the figures have different scales.*

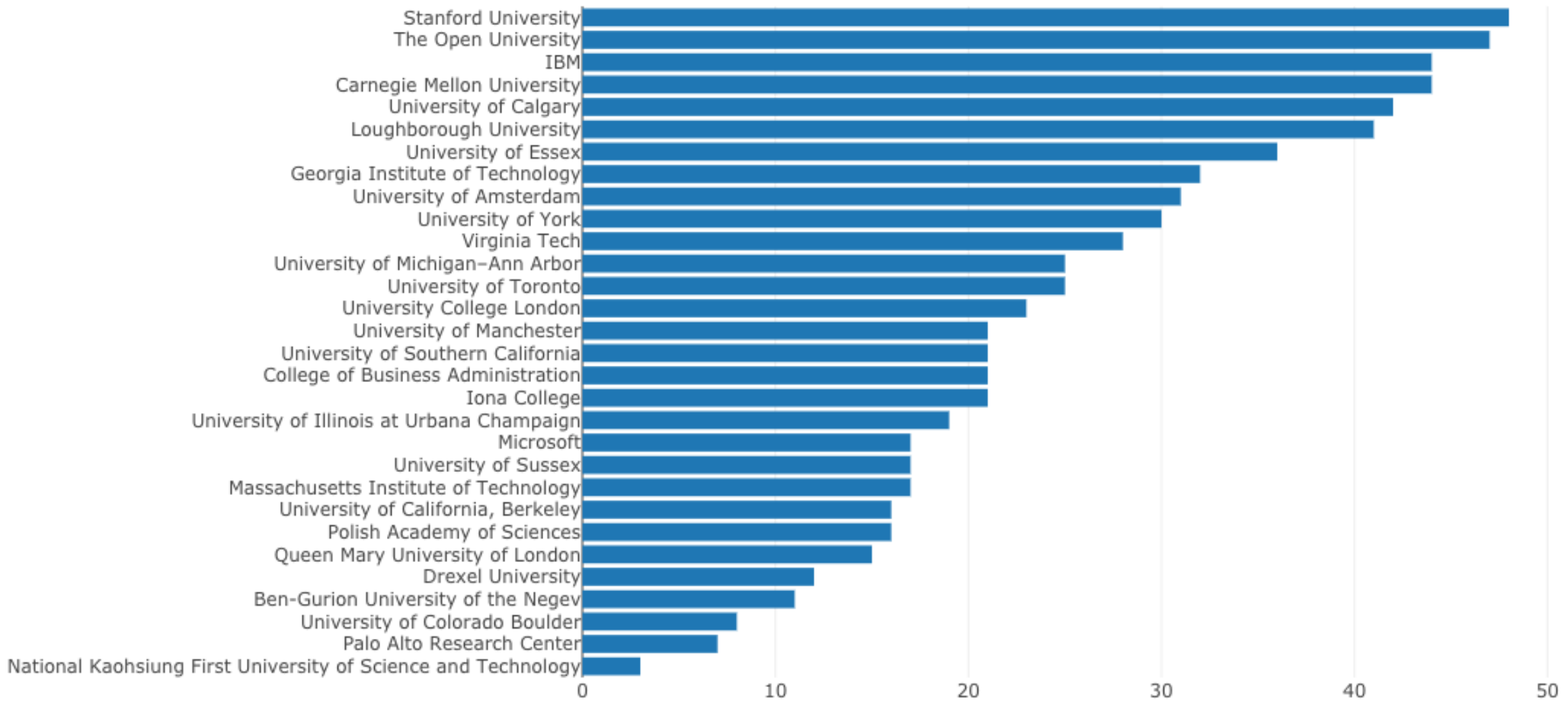

a) Top-30 institutions in IJHCS

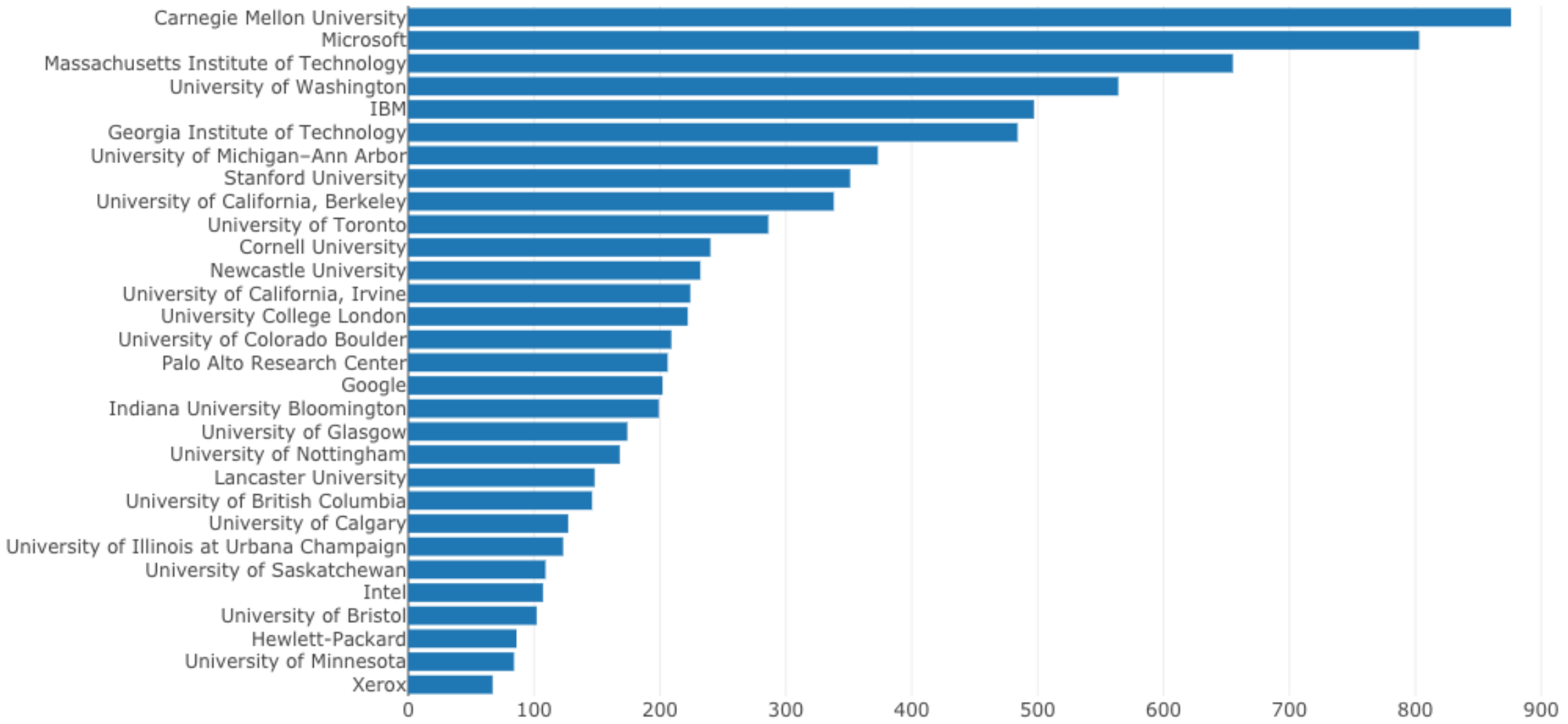

b) Top-30 institutions in CHI

*Figure 2 - Top institutions (a, b) for IJHCS and CHI respectively, ranked by number of publications. When comparing IJHCS and CHI, please consider that the figures have different scales.*

Figure 2 shows the institutions with most publications in IJHCS (a) and CHI (b). The top 10 places in IJHCS include 4 institutions from the US, 4 from UK, 1 from Canada, and one from The Netherlands. CHI has instead a strong North-American profile with 9 of the top 10 places taken by US institutions[12] and the last one by the University of Toronto.

Figure 3 reports several results describing the cited and citing venues. For the sake of completeness, we also include the number of unspecified venues with the indication ‘’n/a”. Figures 3a and 3b show the top-30 most cited venues by IJHCS and CHI respectively. Consistently with the interdisciplinary nature of IJHCS, we can see that its top cited venues include both CHI and IJHCS, as well as other main computing and psychology journals. Conversely, CHI has a stronger focus on HCI venues, citing mainly itself and SIGCHI-related venues (9 out of the top-30 venues).

[12] However, we should point out that the top ten list includes also IBM and Microsoft, which have research labs in various parts of the world.

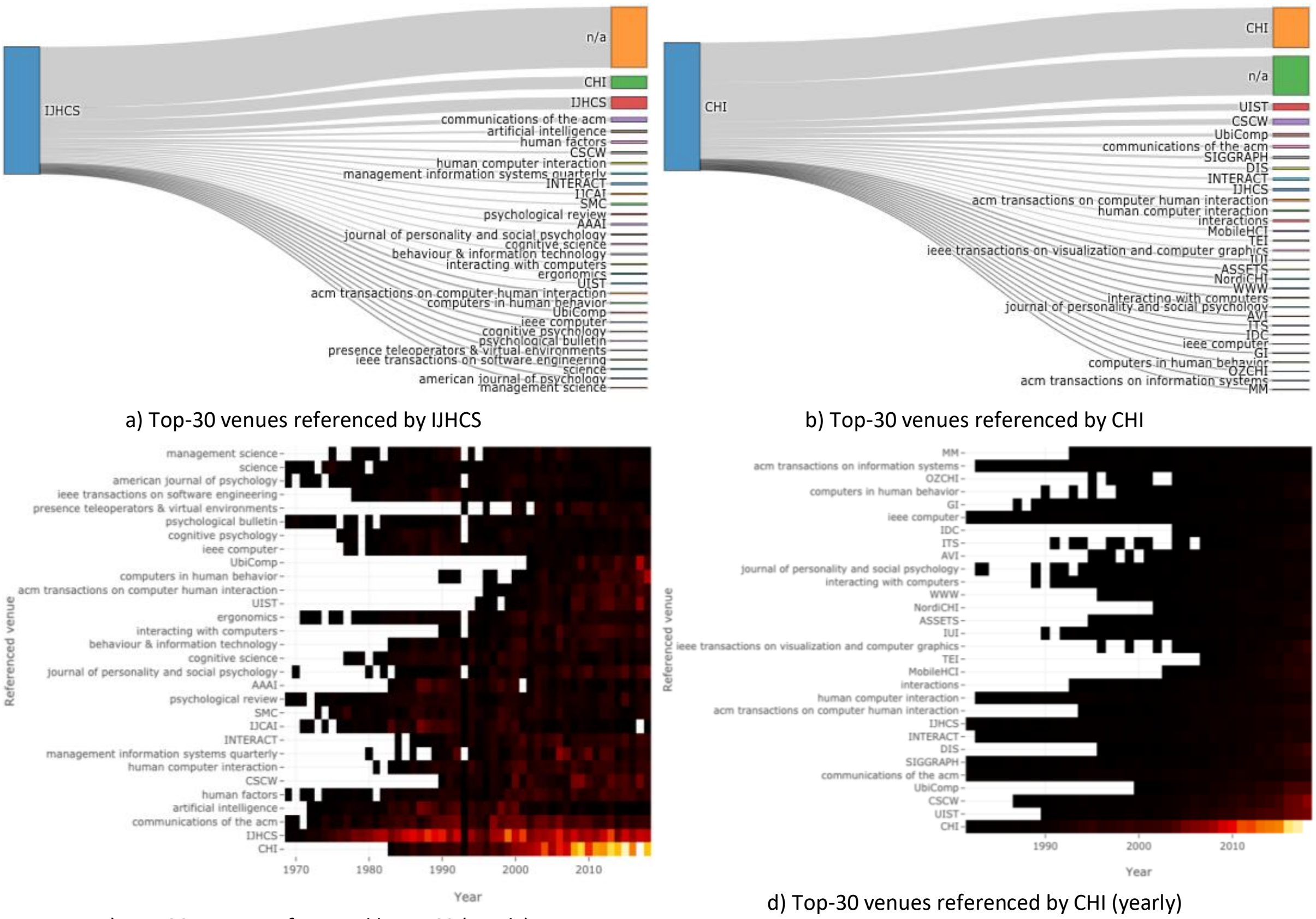

a) Top-30 venues referenced by IJHCS

b) Top-30 venues referenced by CHI

c) Top-30 venues referenced by IJHCS (yearly)

d) Top-30 venues referenced by CHI (yearly)

*Figure 3 - Top-30 venues citing IJHCS and CHI (e, f) and the yearly breakdown of the citations received by IJHCS and CHI, respectively, (g, h) from them.*

The two heatmaps in Figures 3c and 3d show the number of papers cited by IJHCS and CHI over the years in the top-30 cited venues. IJHCS cites fairly evenly a wide variety of venues (i.e. a mild transition red-to-black, from bottom to top). Conversely, CHI authors mainly cite articles from CHI itself, UIST, and CSCW (i.e. a sharp red-to-black transition from bottom to top). Figures 3c and 3d also confirm that IJHCS draws its references from a more multidisciplinary pool of venues than CHI. IJHCS also tends to spread its references more evenly across the years (i.e. scattered brighter pixels in the heatmap), while CHI tends to focus much more on recent research papers (i.e. the bottom-right corner is the one with the brighter shades). These findings are intuitively consistent with the traditional difference in nature between journal and conference publications: while the latter are shorter and focus on a specific novel research result, the former tend to describe a longer arc of research and include a more comprehensive literature review.

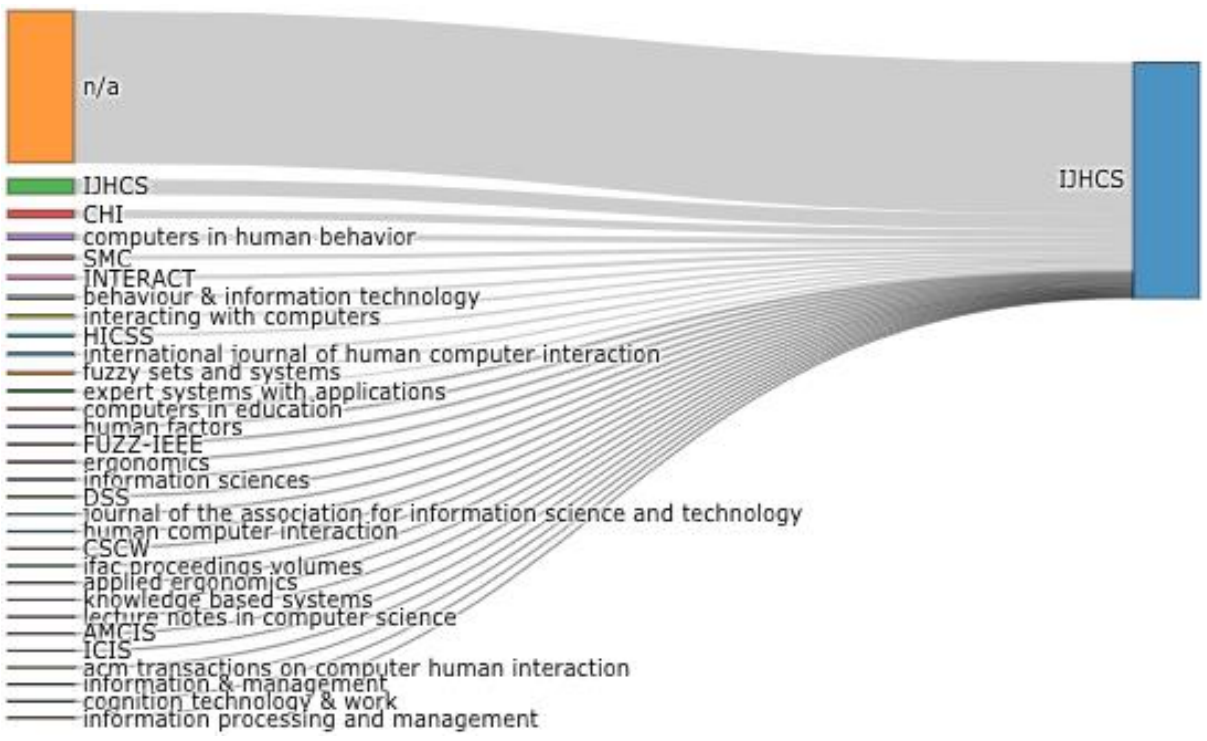

a) Top-30 venues citing IJHCS

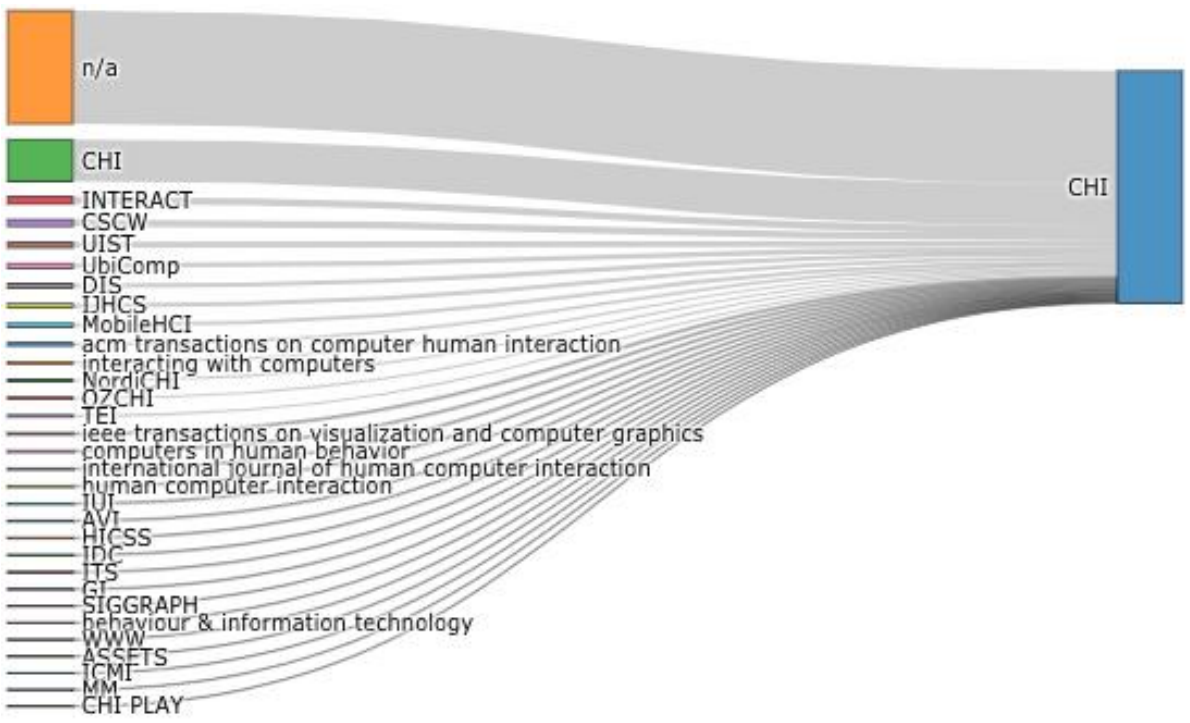

b) Top-30 venues citing CHI

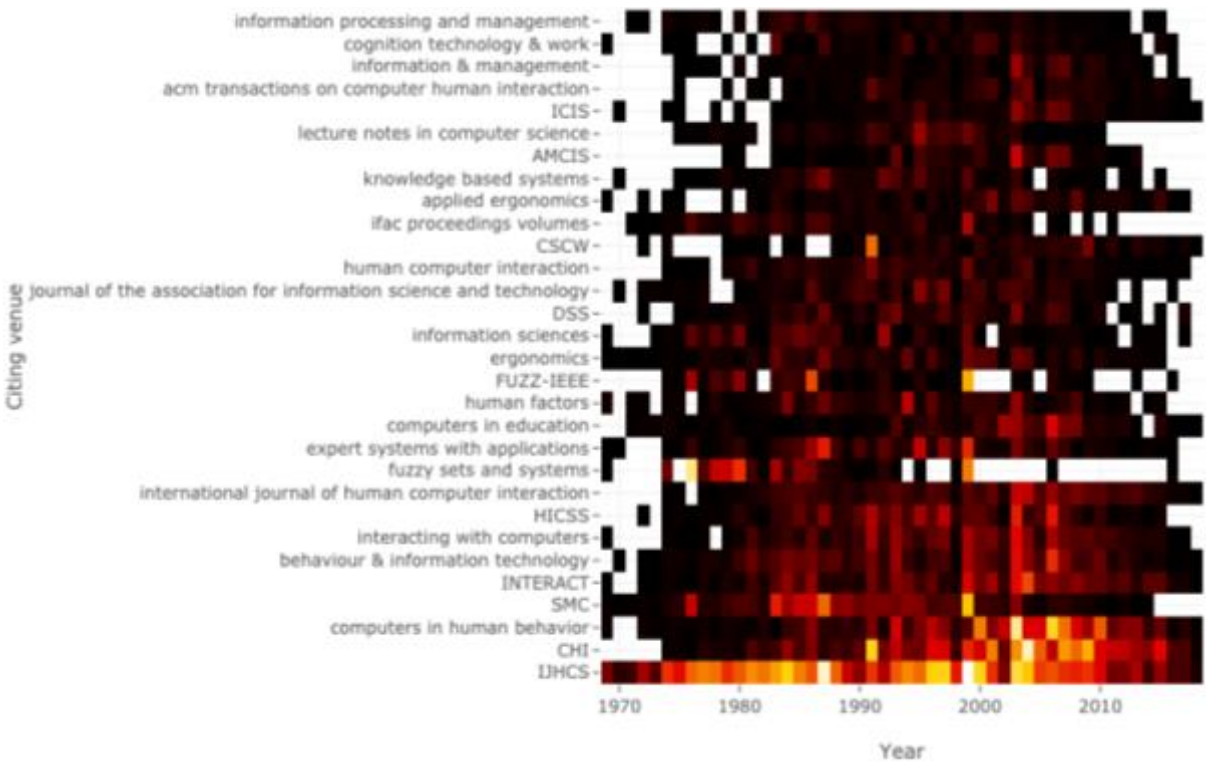

c) Top-30 venues citing IJHCS (yearly)

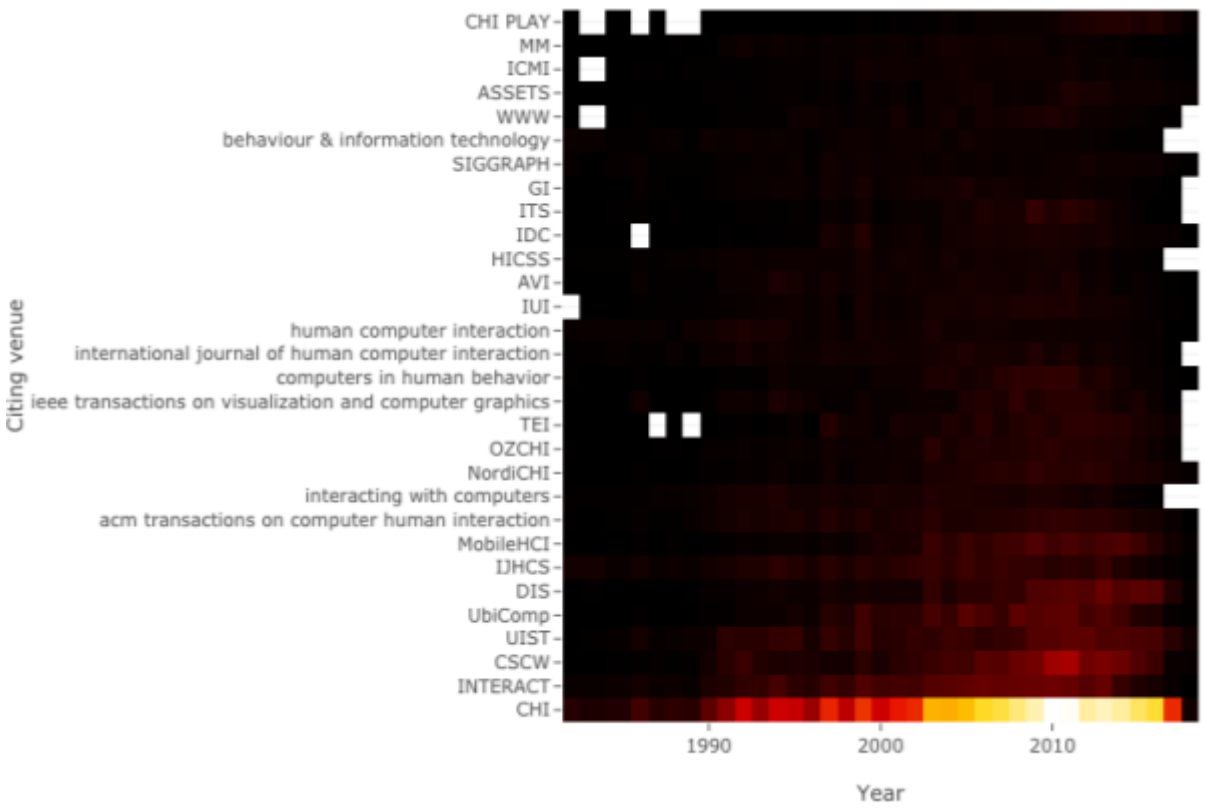

d) Top-30 venues citing CHI (yearly)

*Figure 4* - Top-30 cited venues (a, b) and the yearly breakdown of the citations they received from IJHCS and CHI, respectively (c, d).

In analogy, Figures 4a and 4b show the top-30 venues citing IJHCS and CHI, while the heatmaps in Figures 4c and 4d show the yearly number of citations received by IJHCS and CHI over the years from the top-30 citing venues. Again, the differences are remarkable. The heatmap relative to IJHCS presents scattered bright pixels, suggesting that the journal is capable of producing actionable knowledge for the citing venues regardless of the year in question. Interestingly, the presence of vertical lines of pixels with bright shade indicates that some years were cited very highly across several venues (e.g., 2003). Conversely, papers published in CHI seem to draw attention mainly from CHI and SIGCHI-related venues, as evidenced by the bright bottom row and the fairly uniform red shade located in the bottom right corner (around 2010).

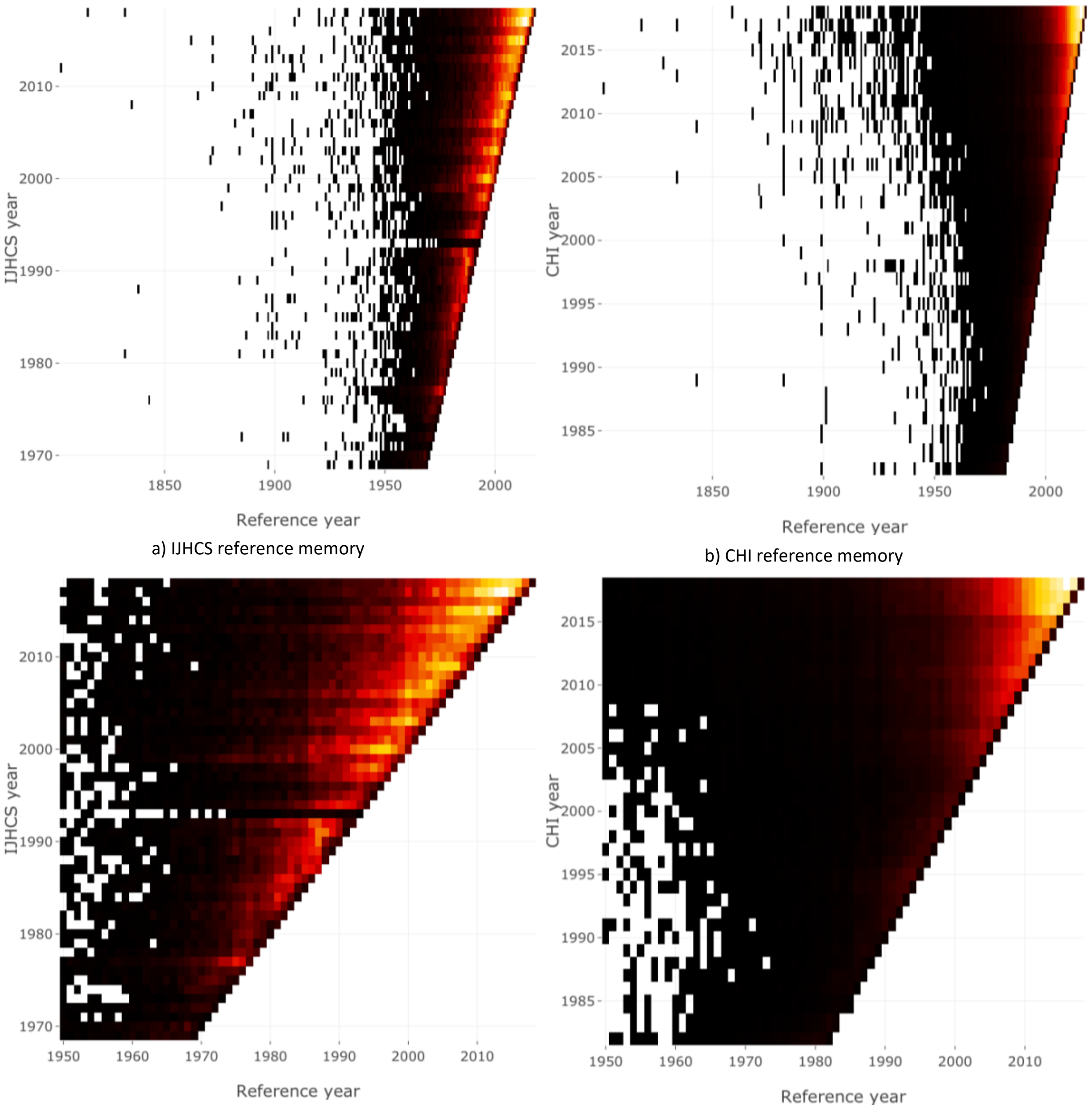

a) IJHCS reference memory

b) CHI reference memory

c) IJHCS reference memory (zoomed since 1950)

d) CHI reference memory (zoomed since 1950)

*Figure 5 – Reference memory: how far back in time the papers cited by IJHCS and CHI (a, b) go. In (c, d) we show a zoomed-in version of this visualization, from 1950 to 2018.*

The heatmaps in Figures 5a and 5b show how old/new are the papers cited by IJHCS and CHI; Figures 5c and 5d show a zoomed version of the same visualisation focusing on the 1950-2018 period. The shade of the pixels encodes (the brighter, the higher) the number of citations given by papers published in the venue in a certain year (on y-axis) to papers of other years (on x-axis). Both IJHCS and CHI draw their references from a broad timespan, including references to papers dating back to early 20$^{th}$ and even the 19$^{th}$ century. Interestingly, IJHCS exhibits a broader attention span to prior work as shown by the evenly bright long tail of pixels for almost each IJHCS year. Besides, IJHCS papers distribute their references fairly evenly among the last 10 to 20 years. Conversely, CHI appears to exhibit a rather short attention span, somehow reflecting the fast-paced nature of computer science conferences, aiming mainly at being on top of the latest developments in the field. This is shown by the shorter and rapidly fading tails of pixels for each CHI year, visible especially in recent years.

In conclusion, it appears that, compared to CHI, IJHCS papers tend to cite a broader range of publication venues across a longer time span. It is tempting to hypothesise that this is an example of a generic phenomenon reflecting the different nature of journal and conference publications, however more research is needed to confirm this.

## 3.2 Geopolitical Analysis

Figure 6 shows the extent to which countries published in (a, b), were cited by (c, d), and cited (e, f) IJHCS and CHI. Figure 7 displays the same data on a histogram plot, revealing the skewed distribution of countries often observed in prior studies [3], [22], [23]. As expected, the least inclusive maps are the ones in Figures 6a and 6b, representing the papers accepted in IJHCS and CHI respectively. Naturally, the pool of papers a venue can refer to or be cited from is broader than the set of papers published in the venue.

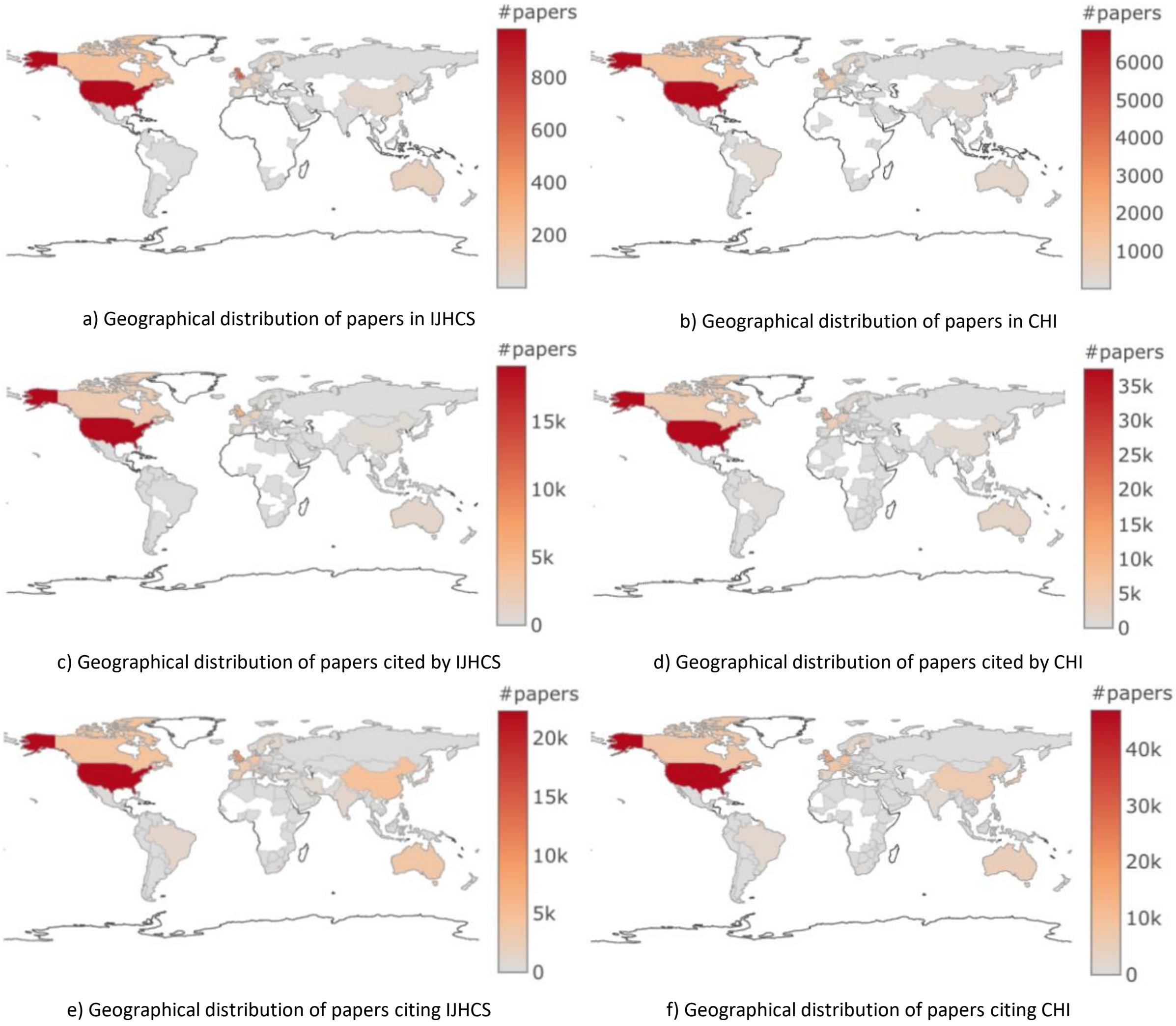

a) Geographical distribution of papers in IJHCS

b) Geographical distribution of papers in CHI

c) Geographical distribution of papers cited by IJHCS

d) Geographical distribution of papers cited by CHI

e) Geographical distribution of papers citing IJHCS

f) Geographical distribution of papers citing CHI

*Figure 6 – Distribution among countries on a world map of publications in IJHCS and CHI (a, b), citations received from IJHCS and CHI papers (c, d) and publications citing papers in IJHCS and CHI (e, f). When comparing IJHCS and CHI, please consider that the figures have different scales. Finally, please notice that countries represented in white are associated to an output of papers equal to zero, while grey shades are associated with values of at least 1.*

Figures 8a and 8b focus on the countries that contributed with five papers or more without collaborating with other countries. USA, UK, Canada, The Netherlands, Australia, Italy, France, Germany and a few other European countries stand out for their ability of working solo in the field. Several countries listed in Figures 6a and 6b are missing here, either because they did not meet the threshold of five papers or because they mainly participated to HCI research through collaborations with other countries. These maps show how difficult it is for countries outside the premier league of research to have a sustained presence in IJHCS and CHI without the support of institutions from the top research countries. Because IJHCS publishes fewer papers per year than CHI, this phenomenon is even more acute in this journal. Here, entire continents, such as South America and Africa, fail to make the cut and, as far as Asia is concerned, only the major far east countries (China, Korea, and Japan) appear on the map.

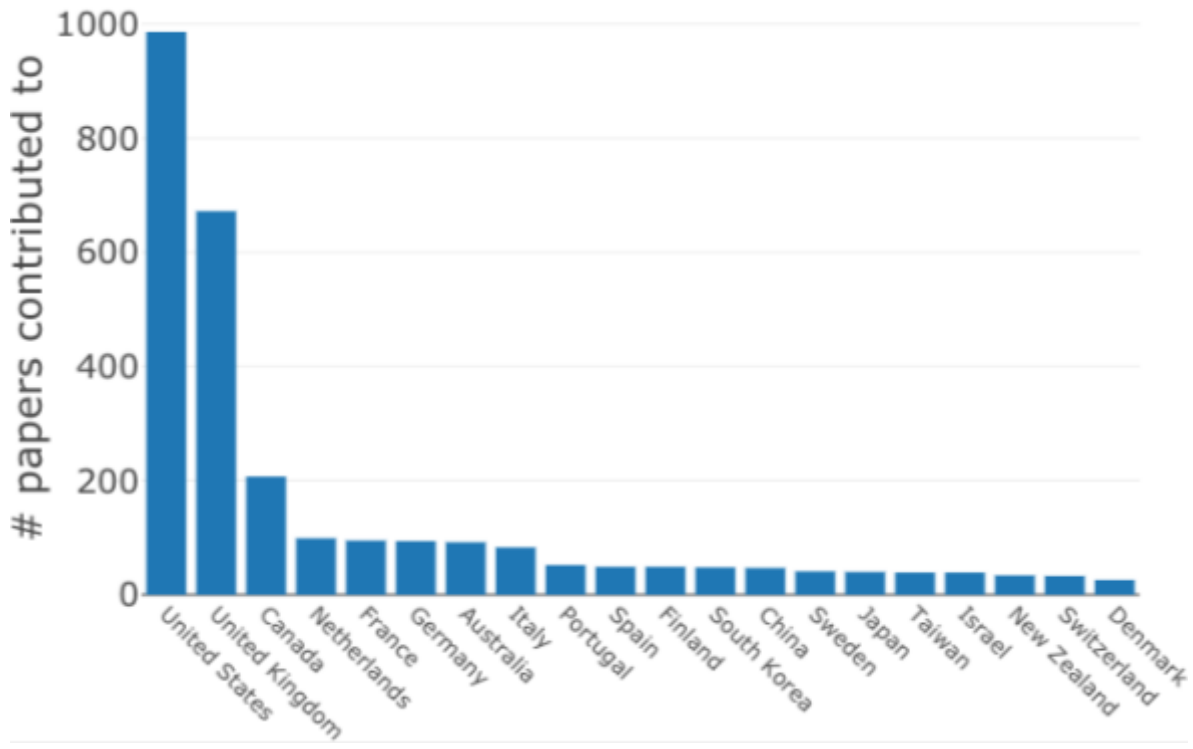

a) Geographical distribution of papers in IJHCS (top-20 countries)

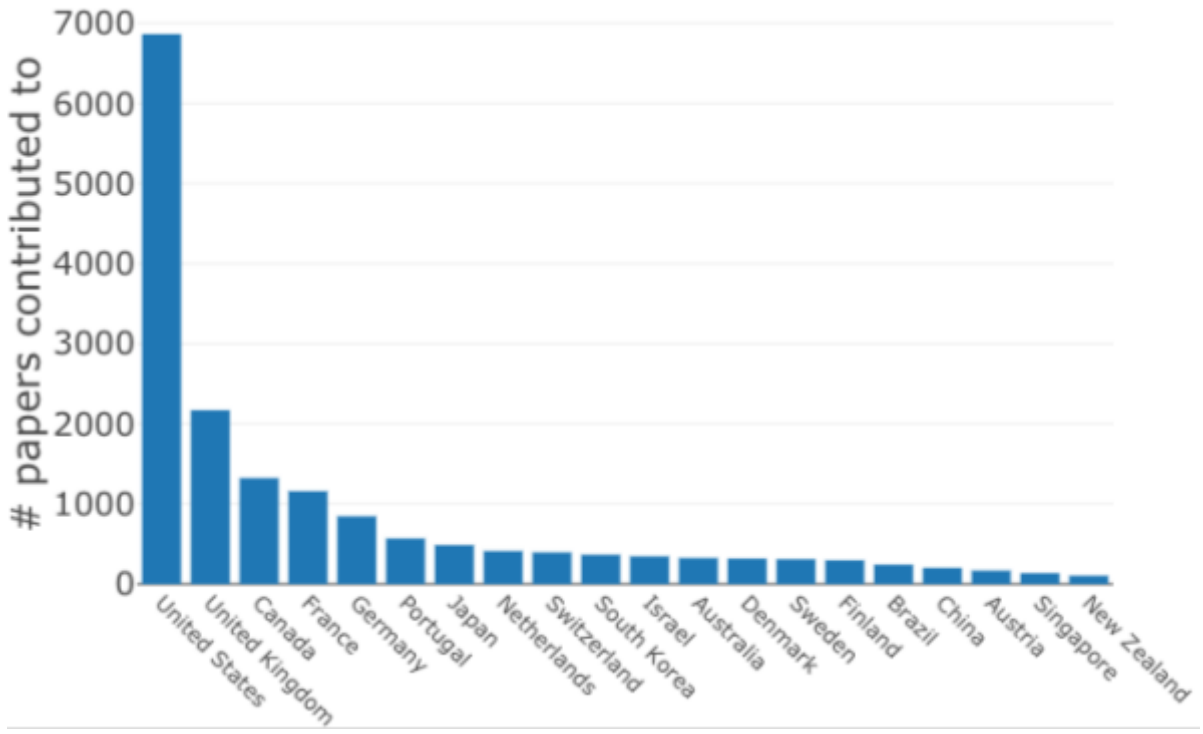

b) Geographical distribution of papers in CHI (top-20 countries)

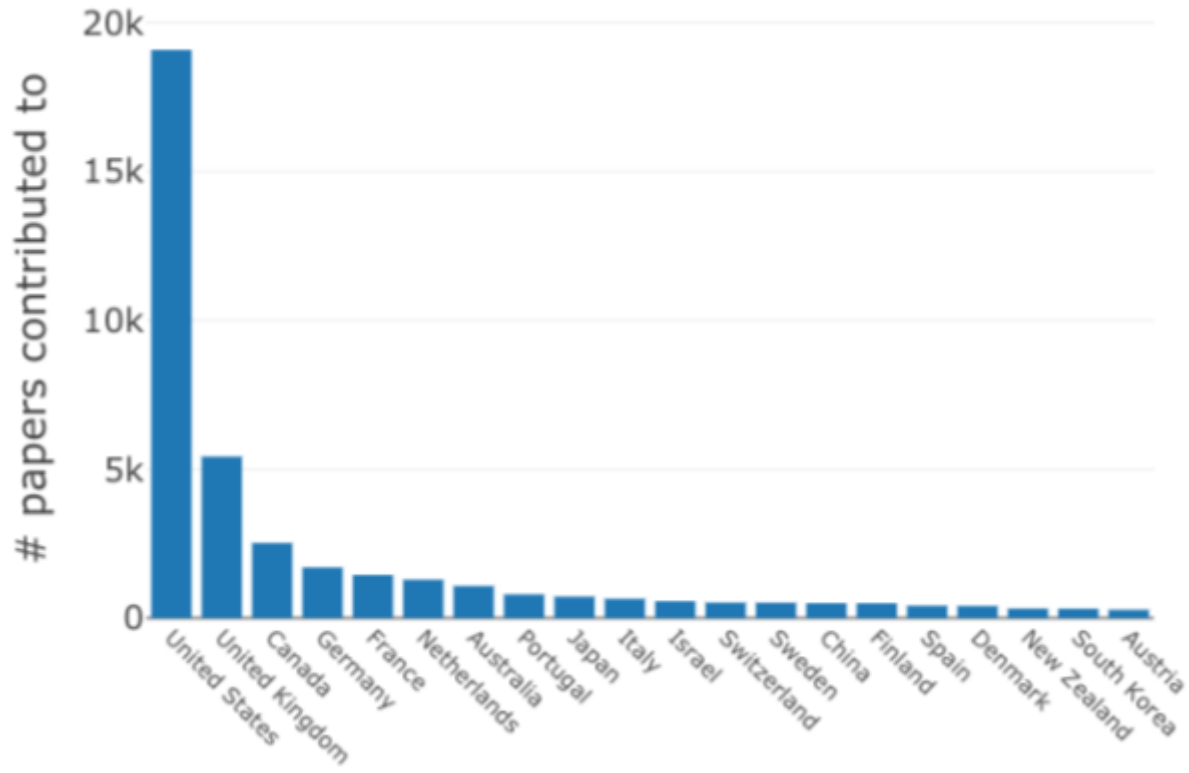

c) Geographical distribution of papers cited by IJHCS (top-20 countries)

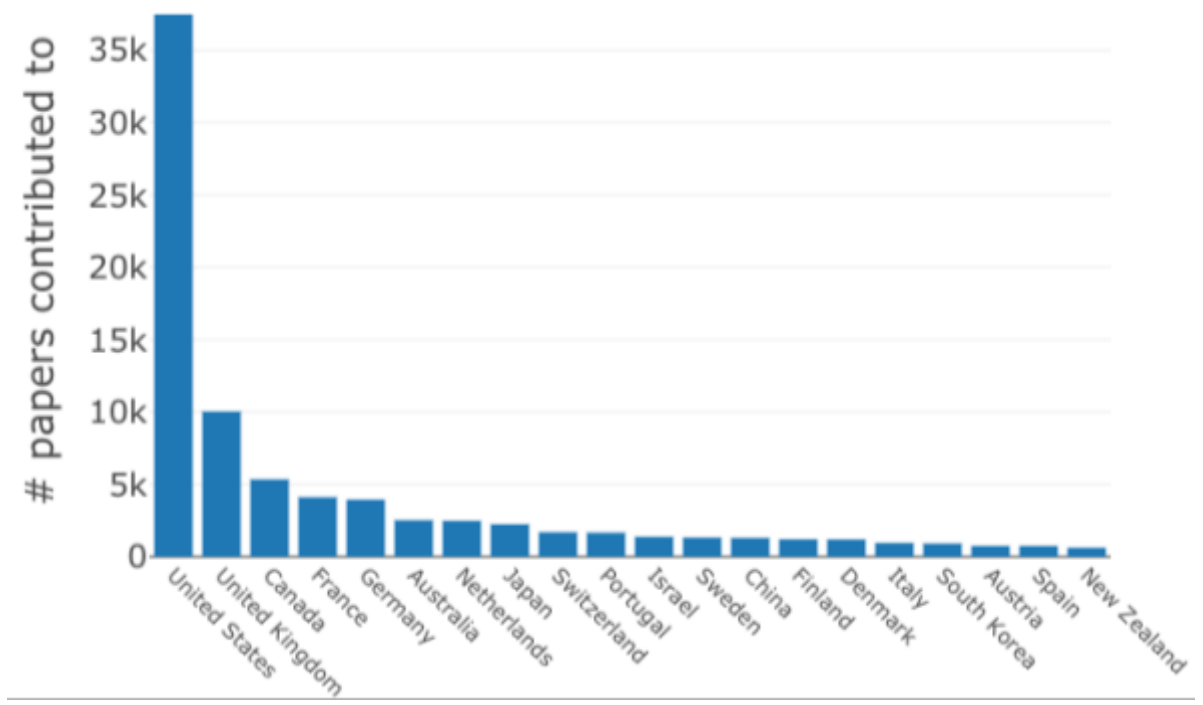

d) Geographical distribution of papers cited by CHI (top-20 countries)

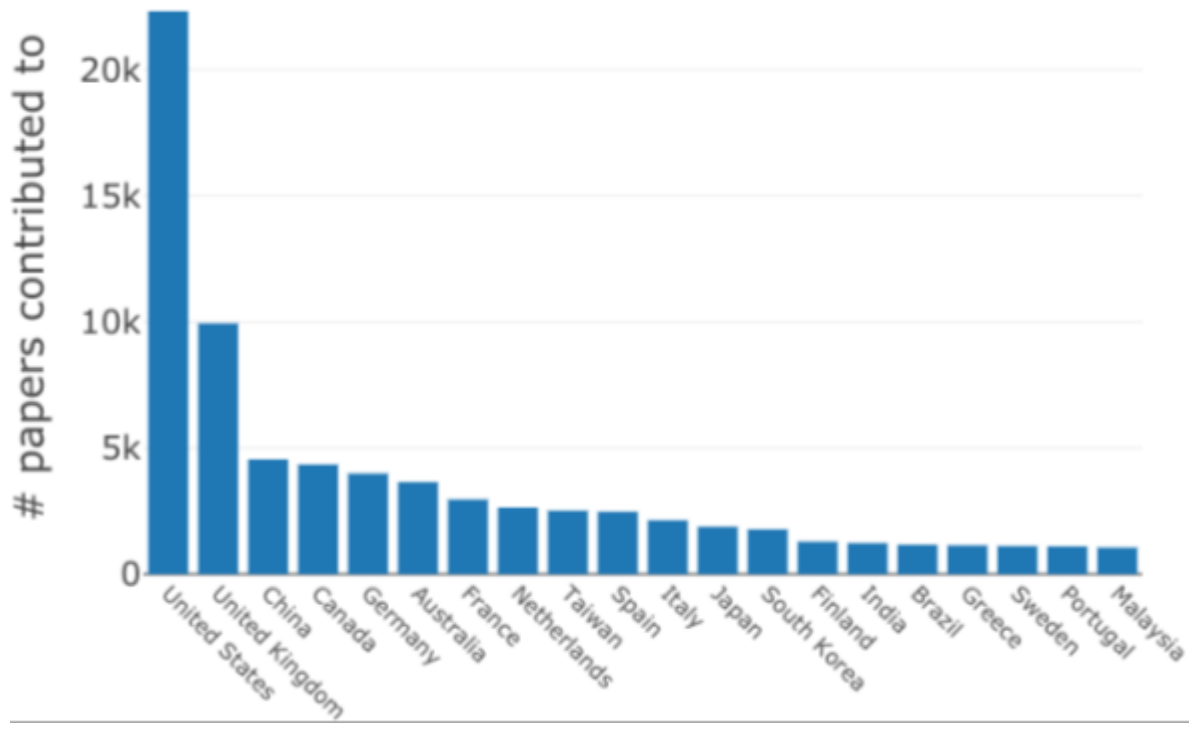

e) Geographical distribution of papers citing IJHCS (top-20 countries)

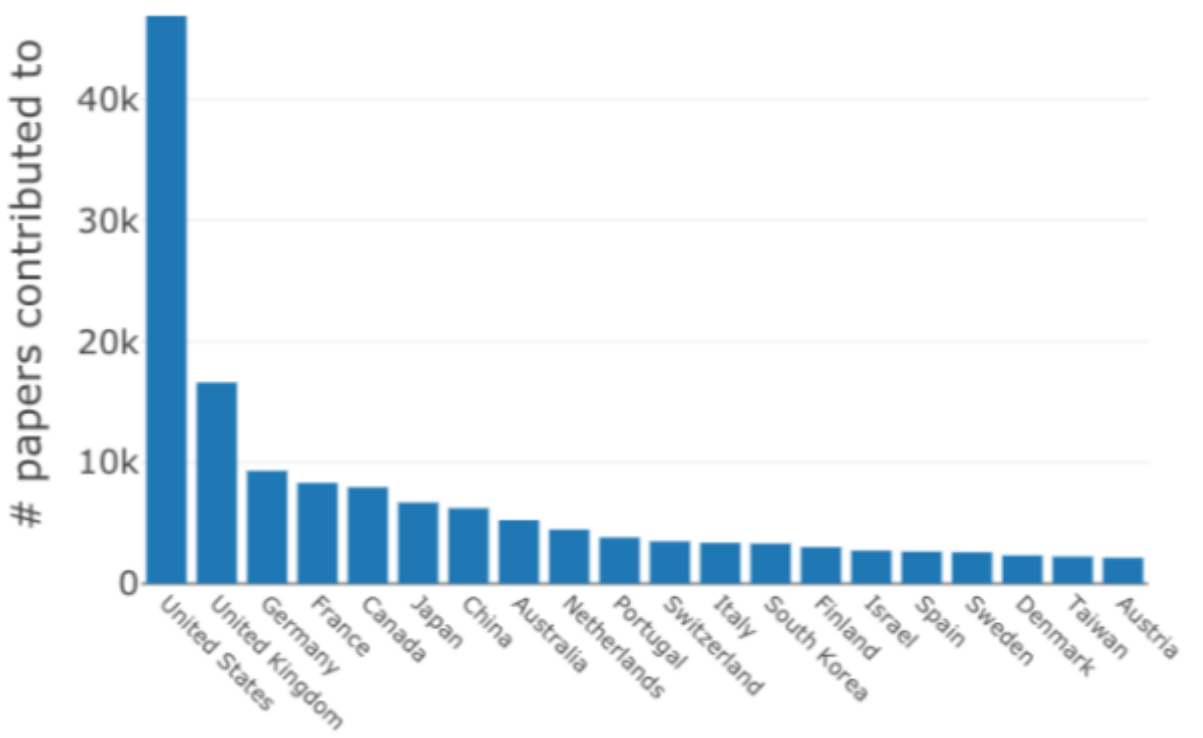

f) Geographical distribution of papers citing CHI (top-20 countries)

*Figure 7 - Distribution among countries of published papers they contributed to (a, b), cited papers (c, d) and citing papers (e, f) in IJHCS and CHI respectively. When comparing IJHCS and CHI, please consider that the figures have different scales.*

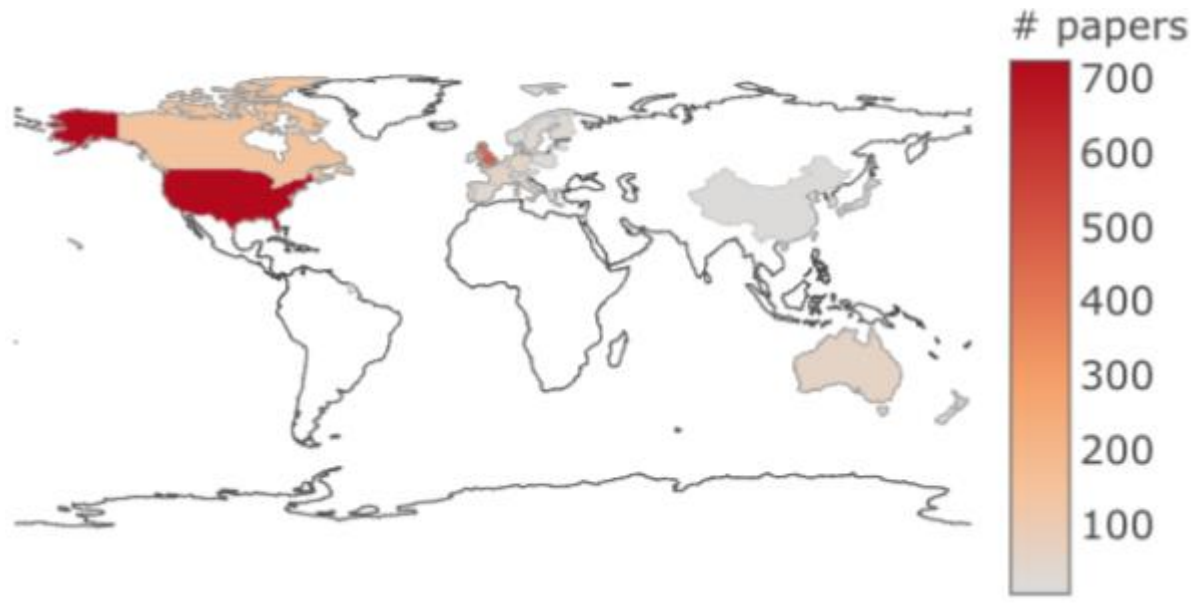

a) Self-sustaining countries in IJHCS

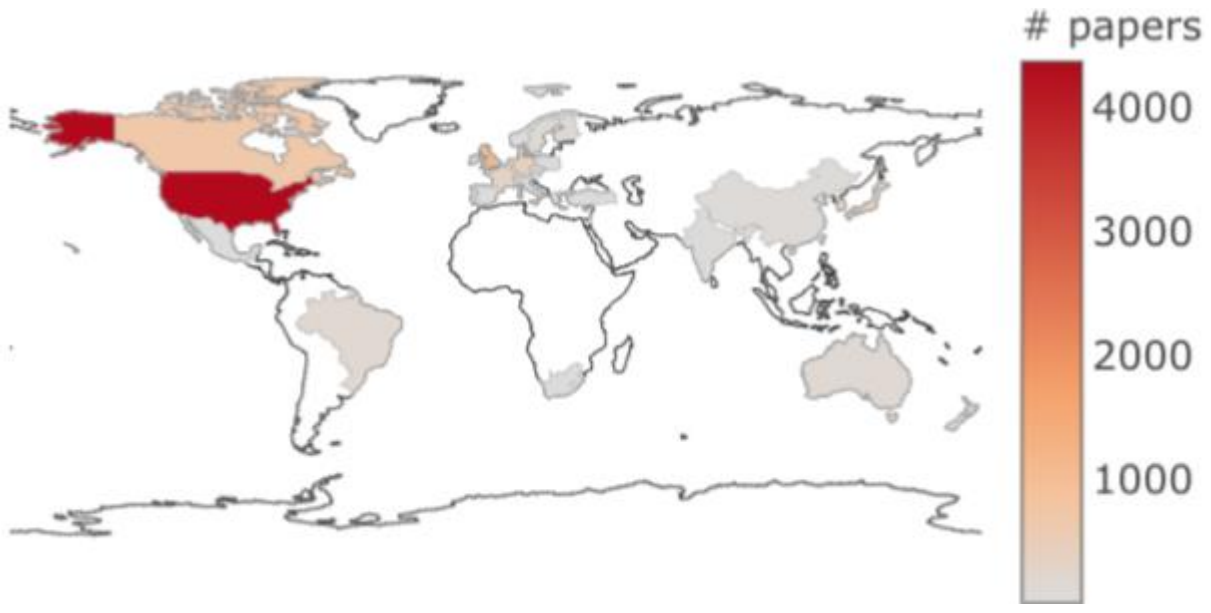

b) Self-sustaining countries in CHI

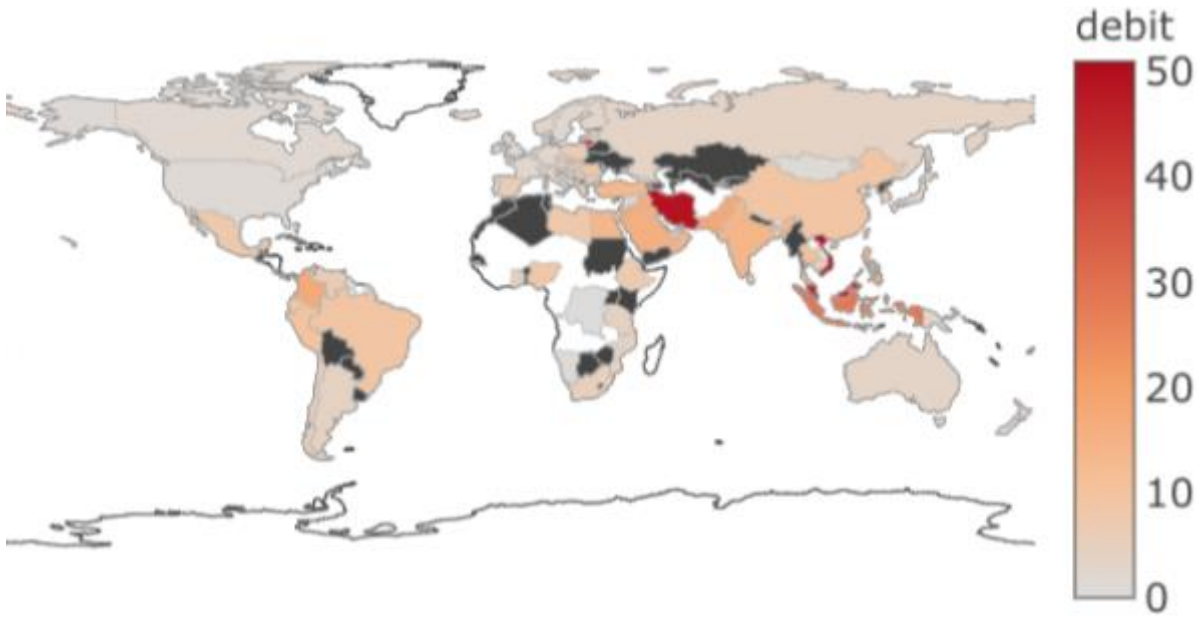

c) Knowledge debit in IJHCS

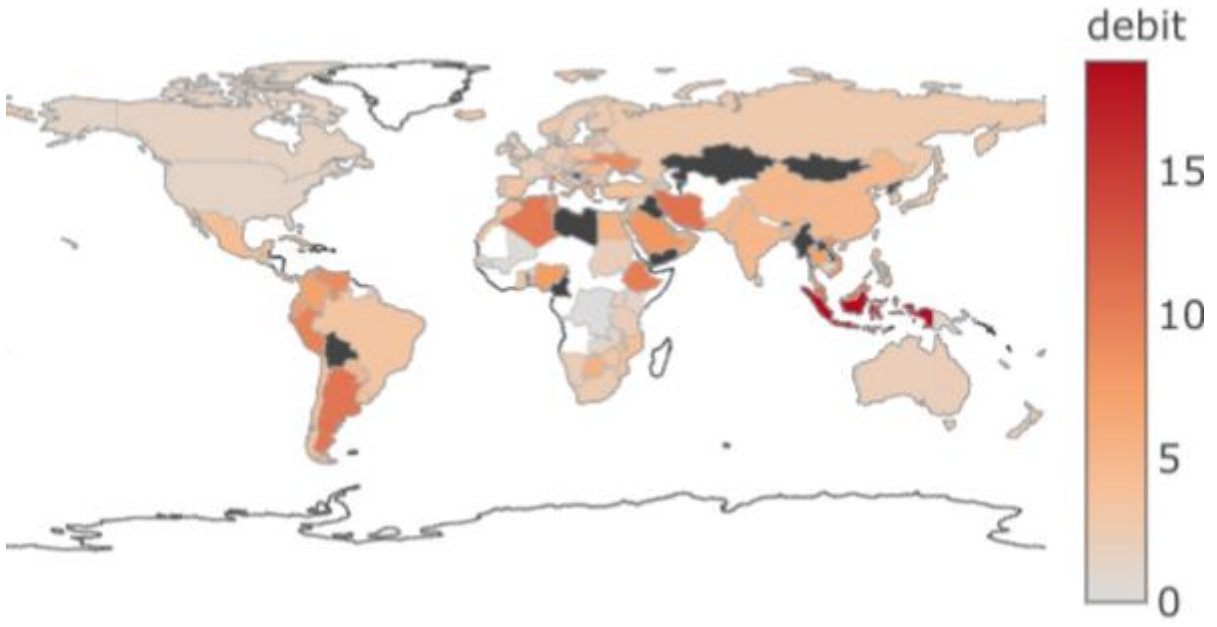

d) Knowledge debit in CHI

*Figure 8 – Paper distribution for countries that managed to publish at least five papers in IJHCS and CHI (a, b) respectively without external collaborations. In (c, d) the knowledge debit of countries as introduced in the methodology section (all papers considered): a country that cites the venues more than it is cited by them appears to be in debit (the brighter, the higher). In black are shown the countries that cite the venues, while being never cited back. Please notice that countries represented in white are associated to a value equal to zero, while grey shades are associated with values of at least 1. Finally, when comparing IJHCS and CHI, please consider that the figures have different scales.*

Figures 8c and 8d show the knowledge debit of countries as defined in Section 2.4 for IJHCS and CHI respectively. In this plot, the brighter the coloration, the higher the knowledge debit towards the venue. The top research countries tend to balance out the amount of knowledge produced for the venue (i.e. citations out) and the amount of knowledge pulled from the venue (i.e. citations in). Conversely, countries such as Iran, India, China, South-Eastern Asian countries, and South American countries cite IJHCS and CHI far more than they are cited by them. The countries that cite the venues but are never cited are instead coloured in black.

These figures show clearly that, at least in the research areas associated with IJHCS and CHI, knowledge generation is confined to a rather small number of countries in North America, Europe, Oceania, and the Far East. Hence, although the number of published papers in the wider international literature has grown dramatically in recent years, this does not appear to have impacted much on the geographical diversity of contributions to top venues, such as IJHCS and CHI.

This trend is further highlighted in Figure 9, which shows the lack of change in country rankings over time, measured in terms of Spearman's rank correlation coefficient (Spearman's rho) [8]. A high value (> 0.7) of the Spearman's rho would point to a strong correlation of rankings in subsequent years, suggesting a stagnant environment in which new entries struggle to emerge. IJHCS oscillates around 0.6, with a slight overall increase in the last decade, while CHI reached 0.9 during the last few years and this value appears to be constantly increasing.

Figure 10 reports, for both IJHCS (a) and CHI (b), the trend over time of unique institutions participating as affiliations of first author (in orange) versus the trend of unique affiliations never appearing as first authors (in green). Generally speaking the number of institutions participating in both CHI and IJHCS is increasing. However, in particular in the case of IJHCS, the rate of increase in the number of institutions affiliating first authors is clearly lower than the one related to institutions never represented by a first author. This is especially the case for the last 10 years. Figures 10c and 10d provide an alternative view by showing the percentage growth of institutions not affiliating first authors for IJHCS and CHI. The number of these institutions is growing in both venues, more so in IJHCS than CHI. This may indicate that an increasing number of institutions is able to participate in IJHCS and CHI only in a (potentially substantial) supporting role.

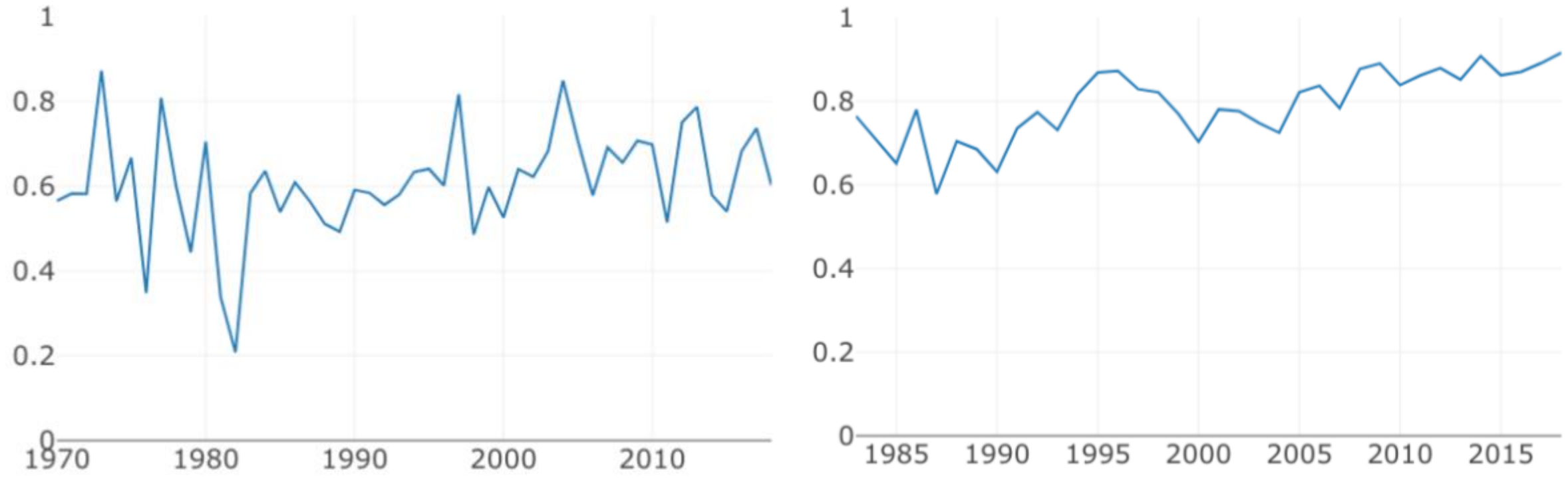

a) Spearman similarity of country rankings for IJHCS

b) Spearman similarity of country rankings for CHI

*Figure 9 – Country ranking similarity over time measured according to Spearman rho.*

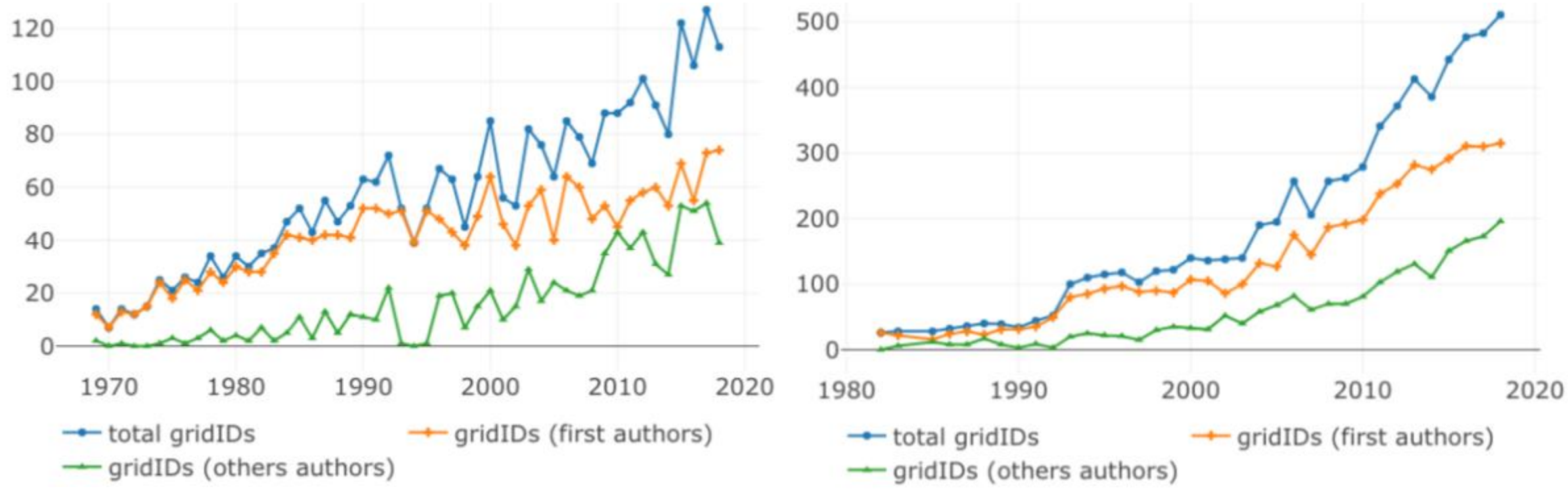

a) GridIds breakdown by author position for IJHCS

b) GridIds breakdown by author position for CHI

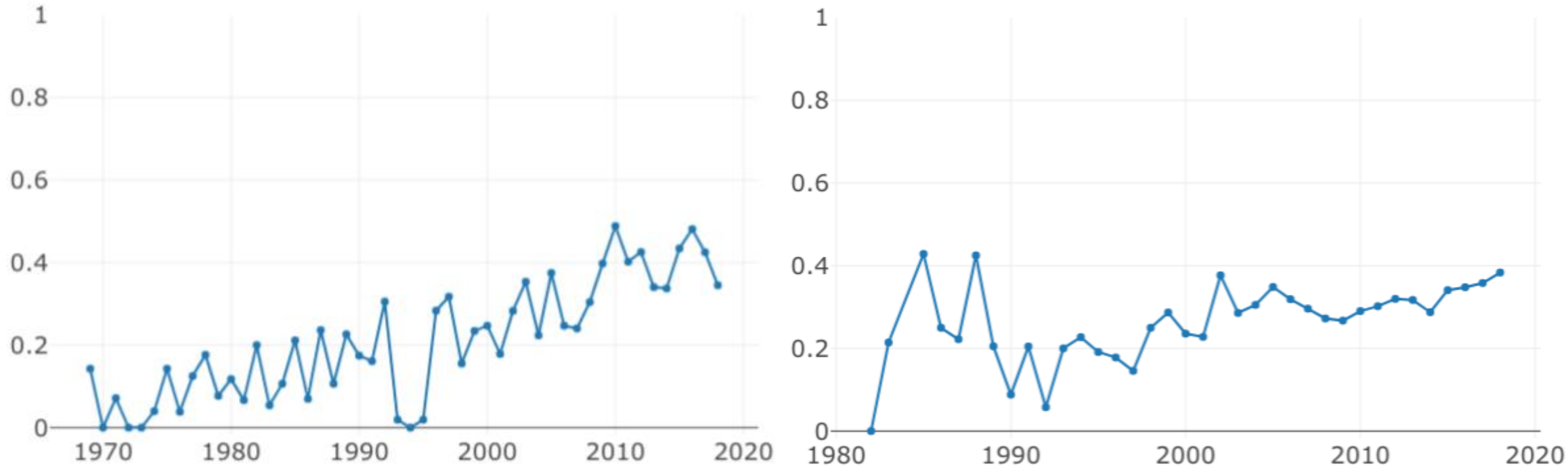

c) Percentage of institutions non affiliating first authors in IJHCS

d) Percentage of institutions non affiliating first authors in CHI

*Figure 10 – Trends of unique institutions affiliating first authors (in orange), unique institutions affiliating authors other than the first ones (in green) and total unique institutions involved in research (in blue) year by year for both IJHCS and CHI (a, b). Percentage of institutions not affiliating first authors (c, d) for both IJHCS and CHI.*

## 3.3 Research Topic Analysis

Figure 11 reports the main topics detected in both IJHCS and CHI and compares the percentage of papers tagged with each topic in IJHCS (blue) and CHI (red). As mentioned in Section 2.5.2, a paper can be tagged with several topics, hence these values do not sum up to 100%. IJHCS and CHI are both prominent venues in the field of HCI and naturally address a similar set of topics. However, as shown in the figure, while CHI is almost entirely focused on HCI, IJHCS has a more interdisciplinary focus, including both HCI and Artificial Intelligence. Hence, topics such as Knowledge Based Systems, Knowledge Management, Formal Languages, and Natural Language Processing play a much more important role in IJHCS than in CHI.

Figure 12 zooms on the first twenty years of IJHCS, reporting on the main topics in the 1969-1988 period, while Figure 13 focuses on the last 10 years, 2009-2018. Interestingly, while historical topics such as Expert Systems are no longer present, it is fair to say that the DNA of the journal has not changed much. Indeed, the key difference is the stronger focus in the last 10 years on HCI and User Interfaces. This, however, has not affected much the other core areas of the journal, with topics such as AI and Knowledge-Based Systems being as important today as they were in the first two decades.

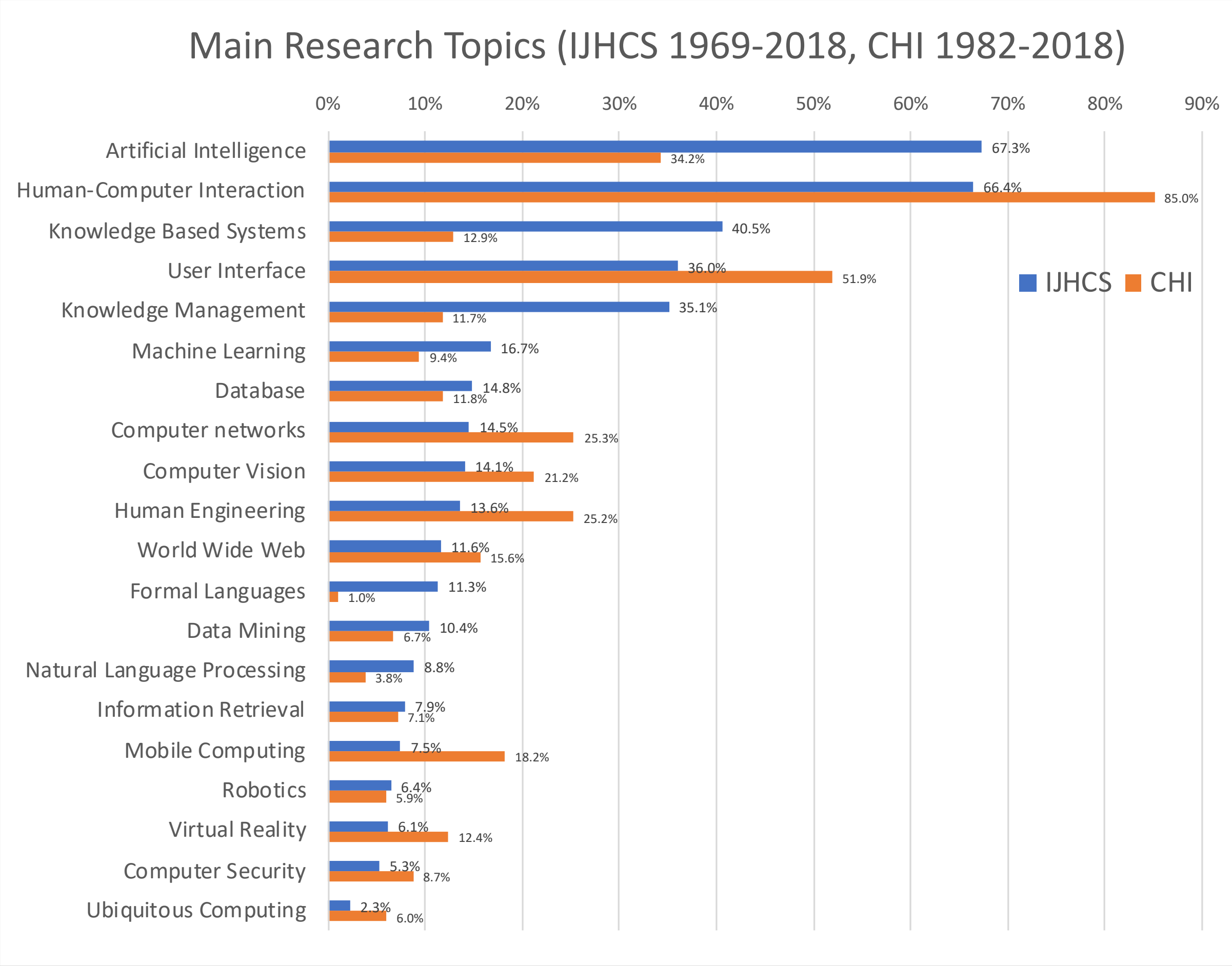

*Figure 11 – Main research topics in IJHCS (blue, 1969-2018) and CHI (red, 1982-2018).*

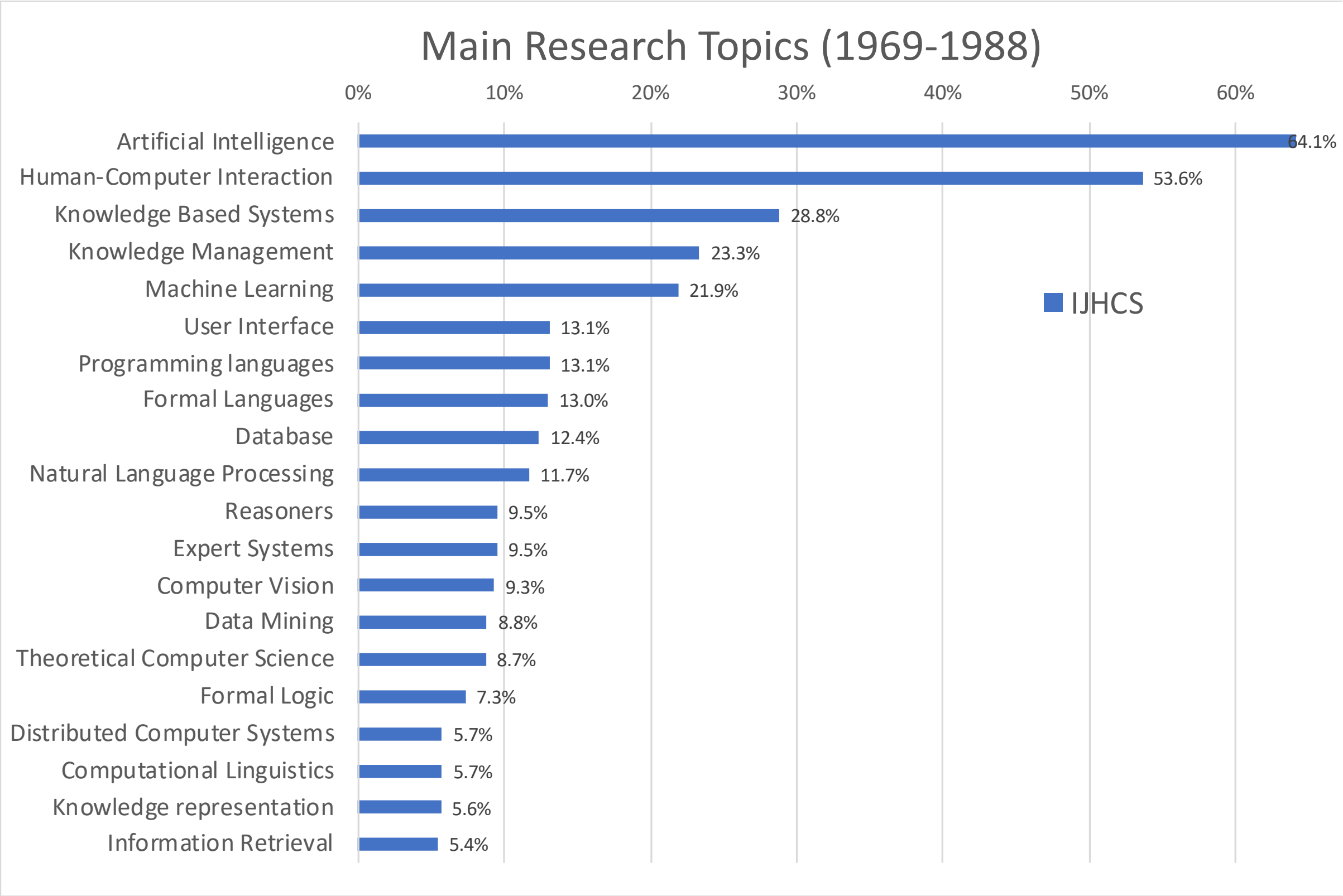

*Figure 12 – Main research topics in IJHCS during the 1969-2018 period.*

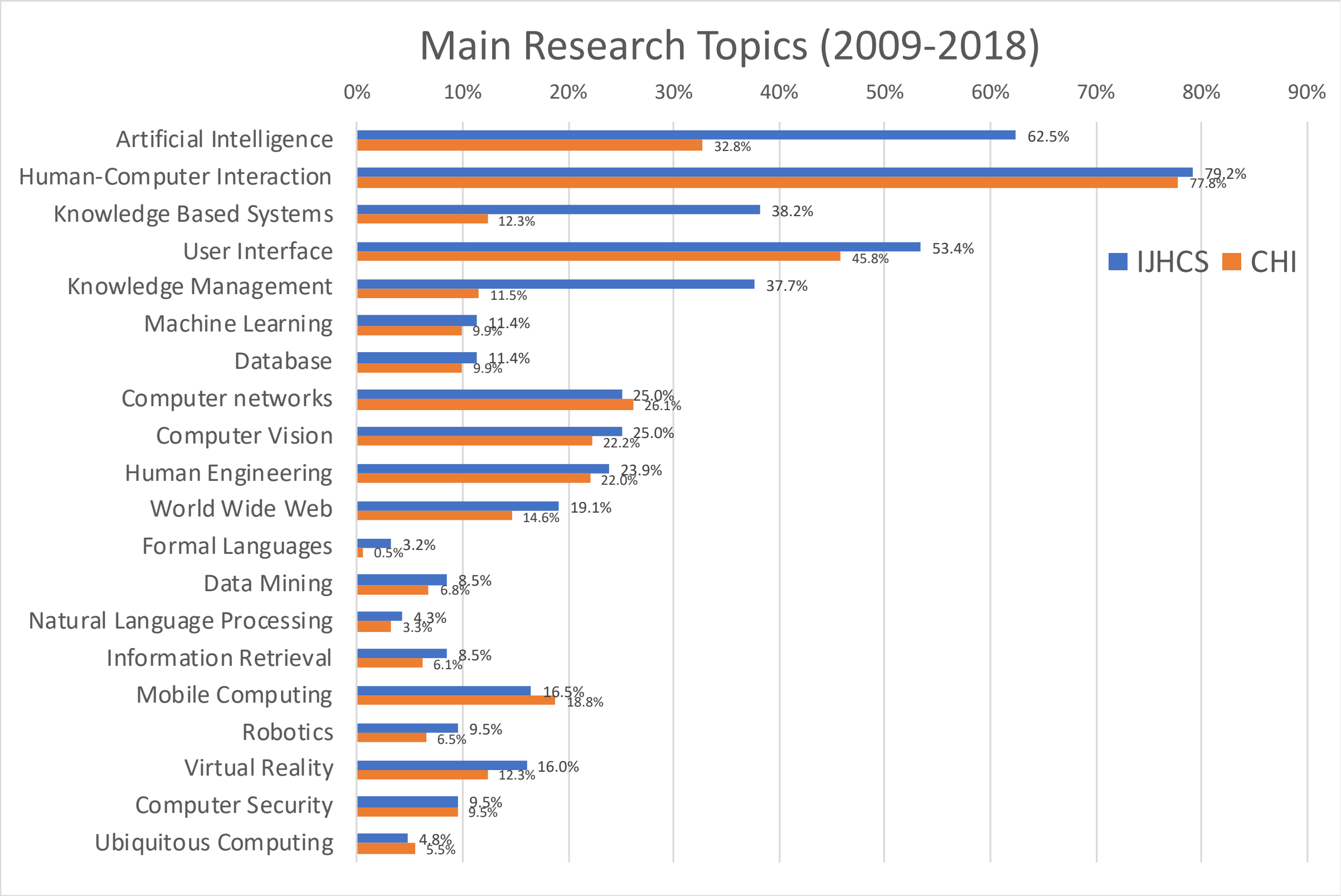

*Figure 13 - Main research topics in IJHCS (blue) and CHI (red) during the 2009-2018 period.*

Figure 13 also indicates that the differences between IJHCS and CHI have become less marked in the last ten years. While IJHCS, as already mentioned, retains a stronger focus on AI, Knowledge-Based Systems and Knowledge Management, the two venues are now very much aligned with respect to the other most prominent topics.

Naturally, the research landscape is very dynamic: new topics emerge continuously, while already existing topics may experience a burst of popularity for a variety of reasons. This can happen because of a breakthrough in the field (e.g., Machine Learning) or the emergence of new consumer products (e.g., Mobile Computing). Figure 14 and Figure 15 highlight the top rising topics in IJHCS and CHI during the 2009-2018 period, grouped according to their number of publications in 2018, as described in Section 2.5.3.

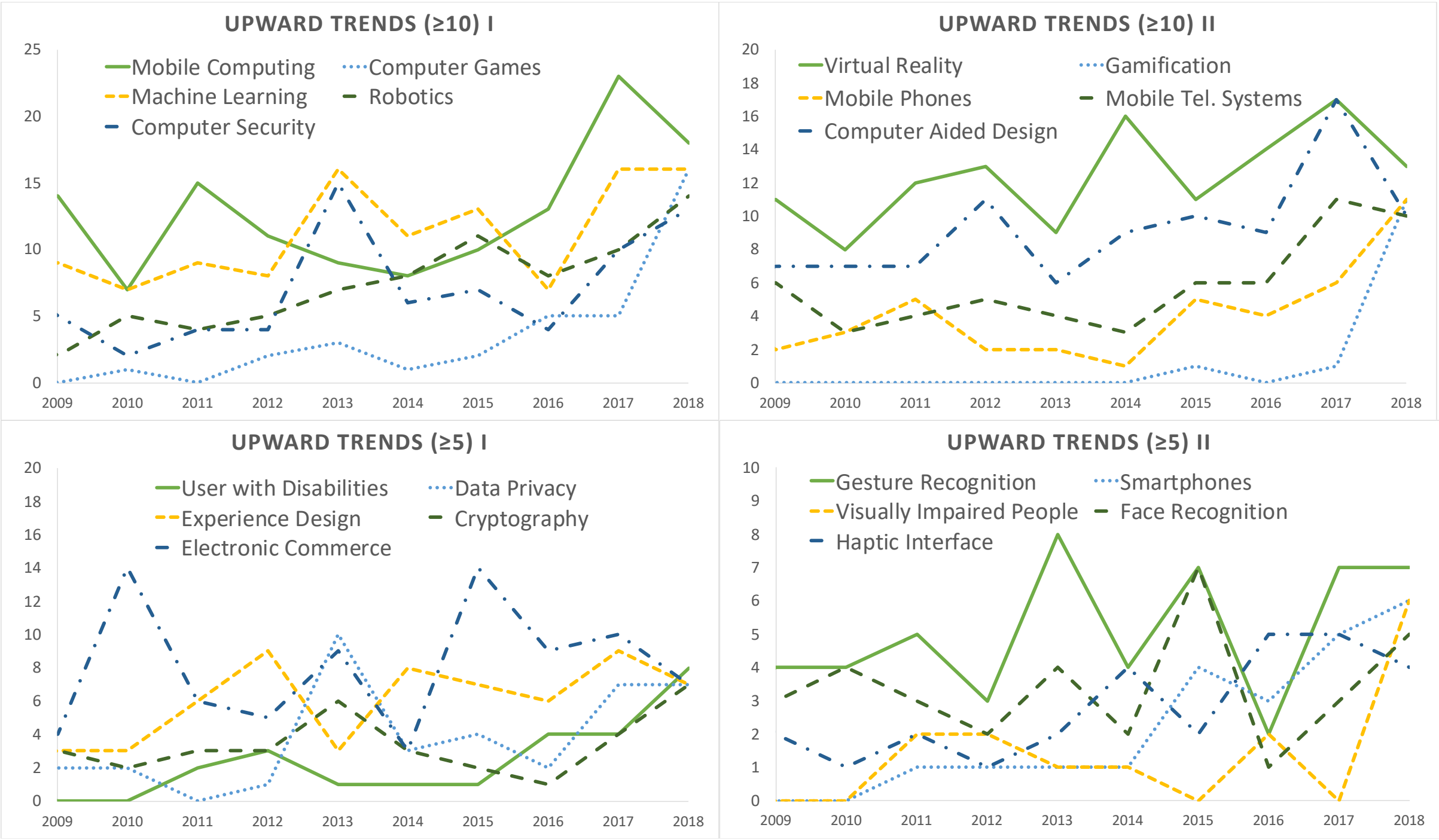

*Figure 14 – Main topic trends in IJHCS during 2009-2018. When comparing IJHCS and CHI, please consider that the figures have different scales.*

Since the large mass of articles published by CHI enables a more fine-grained analysis of emerging topics, in the following we will refer to both venues when talking about coarse-grained topics, but only to CHI when discussing figures for fine-grained topics. Indeed, since the venues are mostly aligned with respect to HCI after 2009, the CHI dataset reflects most of the dynamics involving IJHCS, but also allow us to analyse further the components of the emergent topics.

The most prominent trends emerging in both venues concern Mobile Computing, Machine Learning, Computer Security, Data Privacy, Robotics, Computer Games, and User with Disabilities. Two other additional trends that manifest specifically in CHI regard Virtual Reality and Social Media Analysis.

*Mobile Computing* grew from 142 publications in CHI during 2009 (14 in IJHCS) to 222 in 2018 (18 in IJHCS, with a peak of 23 in 2017). Its most active subtopics in this period include Smartphones (10 to 61 in CHI), Wearable Computing (15, 62), and Mobile Applications (3,19).

*Machine Learning* also experienced a breakthrough in this period, providing solutions to HCI tasks such as face and gesture recognition and rising from 49 (9) to 142 (16). Not surprisingly, the two most active subtopics of Machine Learning during this period are Neural Networks (1 to 22 in CHI) and Deep Learning (0, 10).

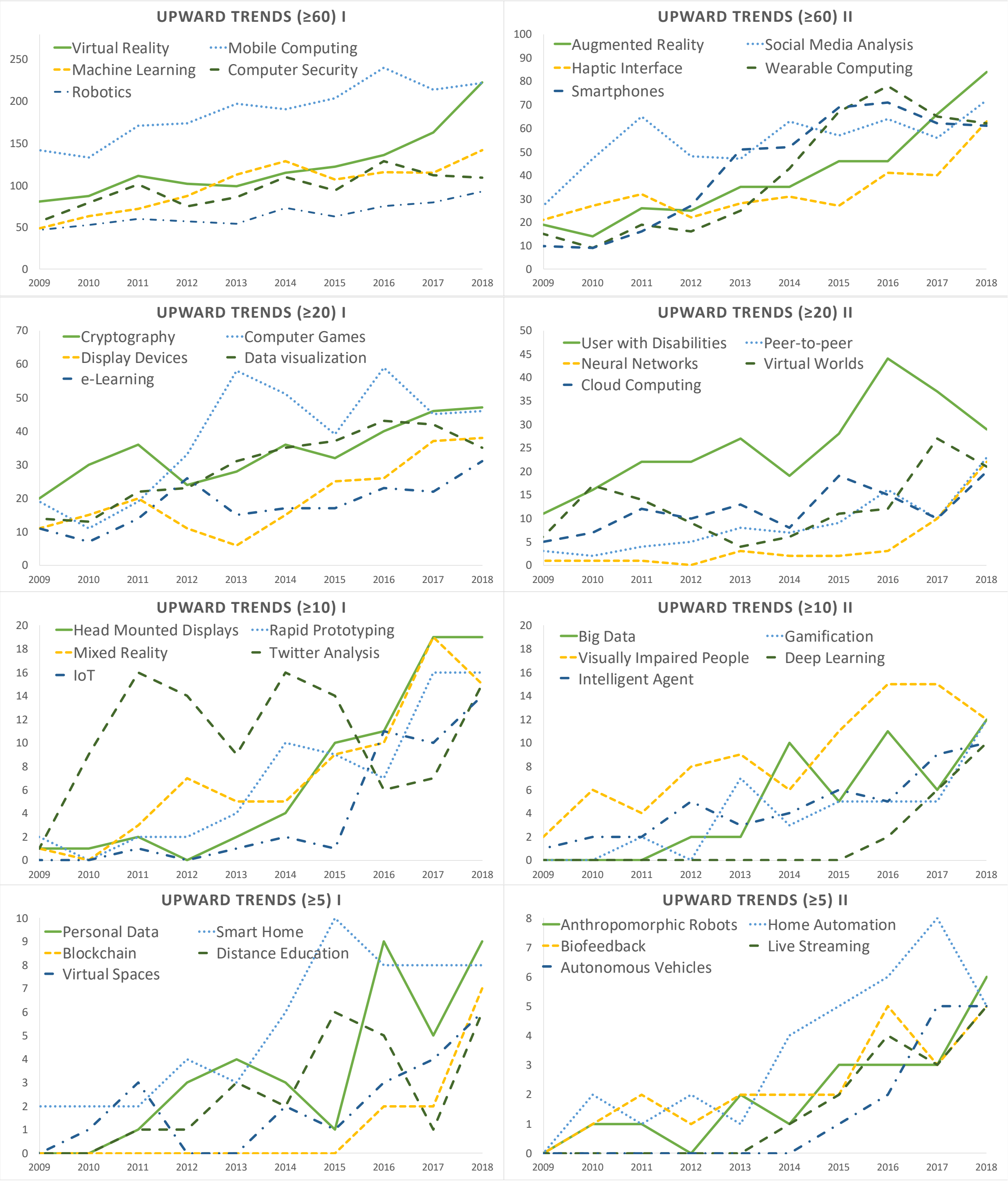

*Figure 15 – Main topic trends in CHI during 2009-2018. When comparing IJHCS and CHI, please consider that the figures have different scales.*

The areas of *Computer Security* and *Data Privacy* have also attracted a lot of attention, growing from 57 (5) to 109 (13) and from 36 (2) to 66 (7) publications. This dynamic is reflected by related topics, such as Cryptography (20 to 47 in CHI), Authentication Systems (14, 26), Privacy Concerns (6, 16), Computer Crime (5, 14), and Personal Data (0, 9).

In the same period, the field of *Robotics* also exhibited a strong growth, going from 47 (2) publication in 2009 to 93 (14) in 2018.

The field of *Computer Games* also experienced a significant acceleration rising from 19 (0) to 46 (16). Key sub-topics here include Gamification (0 to 12 in CHI) and Virtual Words (6, 21).

Finally, an interesting trend regards the area of *User with Disabilities*, which rises from 11 (0) to 29 (8), suggesting a greater sensibility concerning disability issues in both venues. The most active sub-topic in this period was Visually Impaired People (2 to 12 in CHI).

Two other interesting trends manifest mainly in CHI. The first one is the acceleration of *Virtual Reality* which rose from 81 (11) to 223 publications (13 in IJHCS, with a peak of 17 in 2017). The second one concerns the field of *Social Media Analysis* which rose from 27 (0) to 72 (2) publications, mainly thanks to the increasing popularity of microblog services such as Twitter as a data source for the analysis of user behaviour. Indeed, Twitter Analysis rose from 1 to 15 publication in CHI.

Figure 15 shows several additional emerging topics in CHI, including many subtopics of the topics involved in the main trends. These include Augmented Reality (19 to 84 in CHI), Haptic Interfaces (21, 63), e-Learning (11, 31), Cloud Computing (5, 20), Head Mounted Display (1, 19), Virtual Environments (1, 16), Mixed Reality (1, 15), Internet of Things (0, 14), Big Data (0, 12), Affective Computing (3, 12), Smart Home (2, 8), Blockchain (0, 7), Antrophomorphic Robots (0, 6), and others.

In order to easily compare trends across CHI and IJHCS, Figure 16 reports the number of papers tagged with a topic during 2009-2013 and 2014-2018 in IJHCS (respectively bright blue and blue) and CHI (bright green, green) for all the major topics depicted in Figure 14 and Figure 15. It is interesting to notice that while some topics naturally experience a steeper increase in one of the venues, most of them exhibit a similar positive trend in both venues, confirming the substantial alignment between IJHCS and CHI in the past decade. The only exceptions are topics which are scarcely represented in IJHCS, e.g., e-Learning, Neural Networks, and Virtual Words.

Finally, we want to analyse the impact of countries with respect to specific topics. Figure 17 reports the top ten countries for number of publications in HCI, Artificial Intelligence, and the other six previously discussed research topics in IJHCS (blue) and CHI (orange). The shape of the distributions is consistent with the analysis of the overall datasets. USA still dominates, being ranked first in all distributions; Great Britain ranks second in 14 out of 16 distributions, and Canada ranks in the top three in 12 distributions. However, we can also see some interesting differences between the overall distributions and those of specific topics, in particular outside of the top three positions. For instance, Japan is ranked 7th when considering the full CHI dataset (see Figure 7b), but in Robotics is third. Similarly, Italy is ranked 8th overall in IJHCS, but it ranks 4th in Virtual Reality and Computer Games. South Korea, which is ranked 12th and 10th in IJHCS and CHI, is particularly active in Computer Security, rising to 4th and 6th when considering only this research topic.

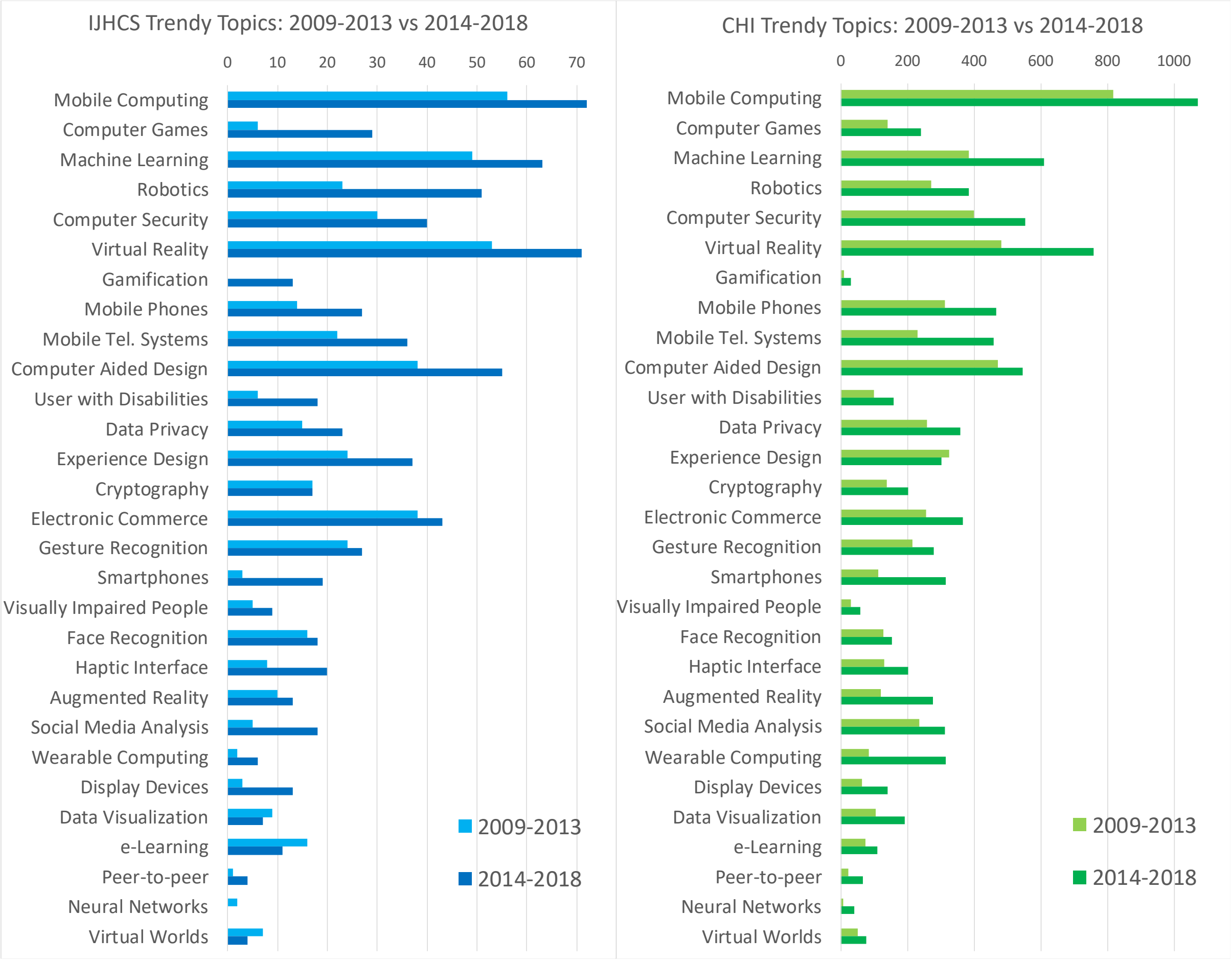

*Figure 16 – Comparison of main research trends in IJHCS and CHI during 2009-2018. When comparing IJHCS and CHI, please consider that the figures have different scales.*

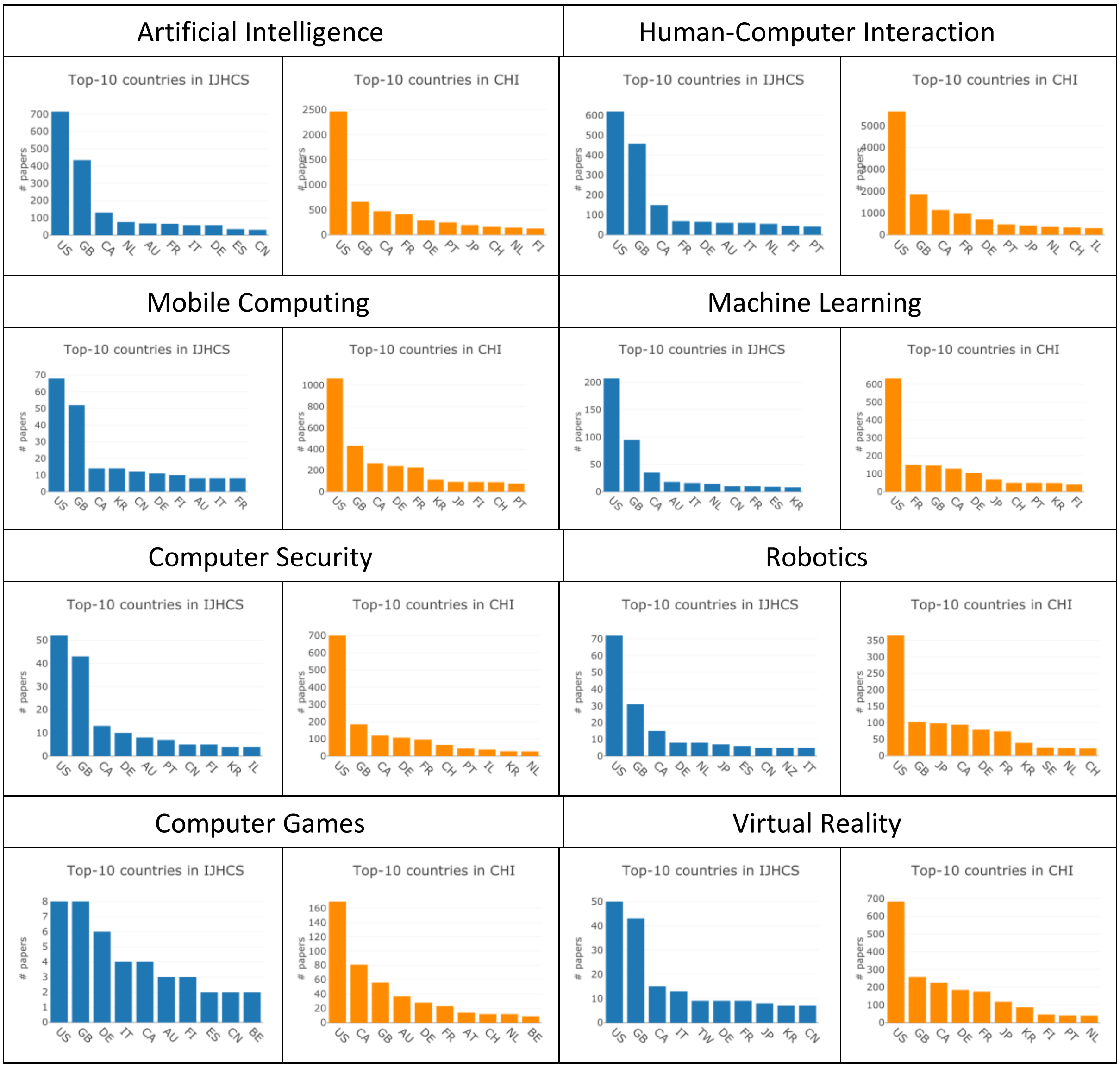

*Figure 17 – Top-ten countries contributing most to eight research topics in IJHCS (blue) and CHI (orange). When comparing IJHCS and CHI, please consider that the figures have different scales.*

# 4 Related Work

The analysis of HCI presented in this paper applies several techniques within the field of Science of Science (SciSci) [1]. The recent emergence of this field has been driven by the increased availability of large amounts of scholarly data, which make it possible to conduct a wide range of analytical studies. For instance, Chen et al. [24] applied CiteSpace to analyse 35,963 articles on Regenerative Medicine between 2000 and 2011. The resulting visual analytics were used in addition to traditional systematic reviews to track the development of new emerging trends. Serenko et al. [22] performed an analysis on 2,175 articles published by 11 journals in the field of Knowledge Management and Intellectual Capital (KM/IC) during the 1994-2008 period. Consistently with our findings, they also report an unbalanced scenario in which few countries produced most of the high-level research in the field. Voracek and Loibl [25] presented a scientometric analysis on publications concerned with studies of the second-to-fourth digit ratio (2D:4D), e.g., correlations with medical conditions or psychological traits, during the period 1998-2008. Their study identified fifteen emerging trends and also found evidence of citation bias. Heilig and Voß [26] analysed 15,376 publications in the field of Cloud Computing during the 2008-2013 period producing several analytics on country distribution and citation patterns. They also performed a statistical analysis of 32,620 unique keywords to identify some emerging topics, such as MapReduce, Data Mining, and Internet of Things.

Similarly to our study, a number of analyses focused specifically on HCI. For instance, Liu et al. [27] analysed the dynamics of 3,152 CHI articles in 1994-2013 using cluster analysis, strategic diagrams, and network analysis. Similarly to us, they identified and discussed the main research trends in the area of Human Computer Interaction by comparing the occurrence of keywords during two time periods (1994-2003 and 2004-2013). Mubin et al. [28] presented a scientometric analysis of the Australian Conference on Human–Computer Interaction during 2006-2015. The main outcomes concern the dominance of a relatively small group of leading researchers and the emergence of research fields such as Design, Health and Well-being, and Education. Marshall et al. [29] analysed over 3,000 citations from 69 papers at CHI2016 and criticised the superficial way researchers talk about previous research in CHI and the practice of "throwaway citations" without critical engagement. Koumaditis and Hussain [30] presented a bibliographic analysis of 962 publications on HCI during 1969-2017, in which they identified a core set of forty-six representative publications, four main thematic areas, and several recent trends concerning workplaces, sensors, and wearables. Chen et al. [31] studied a co-authorship network of 3,620 prominent authors in HCI in the period 1980-2004 and a hybrid network of topical terms and cited articles. The analysis identified seven research areas associated with major trends: Knowledge Representation (KR), WWW, Ubiquitous Computing, Usability Evaluation, User-centred Design, Perceptual Control, and Enterprise Resource Planning (ERP). However, it can be argued that at least two of these areas, KR and ERP, are not really key research areas in HCI. It seems to us that the problem here is caused by conflating HCI (a specific research area) with a number of journals, some of which, in particular IJMMS/IJHCS, are interdisciplinary journals that include not just HCI in their scope, but also a number of other research areas, e.g., knowledge-based systems.

The analysis presented in this paper has elements in common with several of the aforementioned studies, however it arguably provides a more principled and comprehensive analysis of the research dynamics expressed by IJHCS and CHI. In the first instance, we performed a rigorous analysis of the countries and their relationships using the framework presented in [3], exposing a static environment dominated by few countries and resilient to change. In addition, the adoption of the Computer Science Ontology [4] allows us to situate publications more precisely within a particular discipline and detect very fine-grained trends.

The rest of this literature review will focus on spatial scientometrics and approaches to detecting research trends.

## 4.1 Spatial Scientometrics

The field of spatial scientometrics aims at analysing the spatial aspects of the science system. Frenken et al. [5] classified the work in this field in three categories: i) studies that analyse the distribution of publication or citation impact [23, 32–34], ii) studies on the relationship between scientific impact and location [35], [36], and studies that present new approaches to visualise these dynamics [37, 38].

In the first category, we can find several analyses that highlight a great discrepancy in quantity and quality of the research produced by different nations. For instance, May [34] studied the performance of different countries in the 1981-1994 period using the Institute for Scientific Information database, which includes more than 8,4 million papers and 72 million citations. King [33] analysed the same database in the 1993-2002 period and found that the most prominent countries were essentially the same. Pan et al. [23] performed a systematic analysis of citation networks between cities and countries in the 2003-2010 period and reported that the citation distribution of countries and cities followed a power law. Analogously, Huang et al. [32] analysed the Web of Science dataset in the 1981-2008 and discovered that most publications were from a small number of countries. Mannocci et al. [3] examined 506,049 conference papers in 1996-2017 and found that the annual and overall turnover rate of the top countries was extremely low, suggesting a static landscape in which new entries struggle to emerge. Similarly to our analysis, several studies focus on specific research fields. For example, Hung [39] analyses 689 journals in e-Learning and ranks countries according to their ability to cover multiple sub-topics. The analysis by Woodson [40] focuses instead on inequalities in the field of nanomedicine.

The second category of work proposed by Frenken et al. [41] concerns citation impact. For instance, He [35] analysed a dataset of 1,860 papers in the Biomedical field and reported that internationally co-authored papers receive more citations than national collaborations. Similarly, Sin [36] conducted a study on 7,489 papers in the field of Library and Information Science and discovered that those articles that include international collaborations and authors from the top research countries tend to be cited more.

Finally, several works propose applications and techniques to analyse and visualise the spatial aspects of science. For instance, Bornmann and Waltman [37] introduce an approach to generating density maps that highlight regions of scientific excellence. Similarly, Bornmann et al. [38] present an application[13] to visualise institutional performance.

The geographic analysis presented in this paper (Section 3.2) falls mainly in the first two categories presented in this literature review. Unlike the aforementioned analyses, in this study we i) focused on the temporal evolution of countries and institutions in IJHCS and CHI during the 1969-2018 period, ii) studied the dominant position of few countries by assessing the correlation of the country rankings over subsequent years, and iii) examined the affiliations that were never represented by a first author.

## 4.2 Topic Trend Analysis

The analysis of the topic trends presented in this paper follows the EDAM methodology [10], which is a methodology originally developed for reducing the amount of manual work involved in mapping studies. A software engineering mapping study [11] is a method to build a classification scheme and structure a field of interest. Naturally, these methodologies can be applied on any field of science that can be represented in a classification schema. Similarly to EDAM, several other methodologies were developed with the aim of reducing the amount of manual work by the authors. For instance, Octaviano et al. [42] propose a strategy to automate part of the primary study selection activity. Mourão et al. [43] present an assessment of a hybrid search strategy for systematic literature reviews that combines database search and snowballing. Kuhrmann et al. [44] provide recommendations specifically for the general study design, data collection, and study selection procedures. Ros et al. [45] propose a machine learning approach to classifying papers by leveraging human experts, who iteratively validate sets of publications produced by a classifier. Conversely, EDAM does not require experts to manually examine research papers, but only to review a taxonomy of research areas.

[13] Mapping Scientific Excellence, www.excellencemapping.net.

The idea of using ontologies to support mapping studies has been discussed in a few papers but has not actually received much attention. The paper by de Almeida Biolchini et al. [46] introduced the Scientific Research Ontology, a conceptual framework with the aim of fostering the consistency between different studies. However, this resource does not directly assist the extraction of primary studies. Sun et al. [47] discussed the use of ontologies for supporting key activities in mapping studies and presented an experiment in which they automatically classified primary studies by means of COSONT, an ontology of methods for cost estimation. Unfortunately, their approach still requires the manual checking of hundreds of papers and the COSONT ontology is quite simplistic, being a handcrafted list of methods with no hierarchical structure. The main advantage of the EDAM methodology [10], adopted in our study, is that it exploits a completely automatic approach to classifying publications and thus does not require experts to manually review a large number of papers.

# 5 Conclusions

In this work we have analysed the evolution of IJHCS and CHI since their very beginnings, aiming to obtain useful insights about the research dynamics associated with these scientific venues. The results confirm some of the intuitions we have about the difference in nature between journal and conference papers. For example, we showed that IJHCS papers tend to cite a more multidisciplinary variety of venues than CHI and spread their citations more evenly across the years, while CHI papers tend to cite primarily very recent papers. We also showed that these highly selective publication venues tend to be rather closed to newcomers and primarily showcase work from a rather small pool of top research countries. As indicated by our analysis that was centred on the notion of knowledge debit, there are a number of countries that show a very high level of interest in what is happening in IJHCS and HCI but are unable to have a significant publishing presence or citation impact in these outlets. This suggests that specific initiatives ought to be put in place, to widen participation in these scientific venues. Finally, we detected several research trends emerging in the last ten years concerning Mobile Computing, Machine Learning, Computer Security, Data Privacy, Robotics, Computer Games, User with Disabilities, Virtual Reality, and Social Media Analysis.

As this paper was written in the context of celebrating the 50$^{th}$ anniversary of IJHCS, it is also particularly interesting to see that the DNA of the journal has remained rather stable over these 50 years, maintaining a core focus on AI and HCI as the main areas of interest. Considering the difference between the status of AI and HCI today compared to what it was in 1969, this is a remarkable state of affairs, which shows the amazing vision that the founders of IJHCS demonstrated in establishing a journal whose core topics are still regarded as highly important research areas 50 years later. This is even more remarkable if we consider that IJHCS predates the International Joint Conference on Artificial Intelligence (by a few months), the Artificial Intelligence journal (by one year), and CHI (by 13 years).

In sum, the story of IJHCS is one of amazing vision, sustained excellence, and great success. We look forward to following the evolution of the journal in future years.

# 6 Acknowledgements

We thank Microsoft Academic Service for sponsoring our research and enabling us to use their large repositories of scholarly data. We also thank Springer Nature for supporting our research on scholarly analytics over the years.